\documentstyle[11pt,epsf]{article}

\newcommand{\text}[1]{\mathrm{#1}}
\begin{document}

\thispagestyle{empty} \pagestyle{myheadings}
\markright{{\sl Towards single-electron metrology}}

\begin{center}
{\Large {\bf Towards single-electron metrology}} \vskip1cm

Karsten Flensberg$^{*}$\\[0pt]
{\it Danish Institute of Fundamental Metrology,}\\[0pt]
{\it \ Anker Engelunds Vej 1, DK-2800 Lyngby, Denmark}\\[0pt]
and\\[0pt]
Arkadi A. Odintsov$^{**}$\\[0pt]
{\it Nederlands Meetinstituut, P.O. Box 654, 2600 AR Delft, The Netherlands,
and }\\[0pt]
{\it Department of Applied Physics, Delft University of Technology, 2628 CJ
Delft, The Netherlands,}\\[0pt]
and\\[0pt]
Feike Liefrink and Paul Teunissen \\[0pt]
{\it NMi Van Swinden Laboratorium, P.O. Box 654, 2600 AR Delft, The
Netherlands.} \vskip1cm
21 December, 1998.
\end{center}

\vskip1cm

\begin{quote}
We review the status of the understanding of single-electron transport (SET)
devices with respect to their applicability in metrology.
Their envisioned role as the basis of a high-precision electrical standard is
outlined and is discussed in the context of other standards.
The operation principles of single electron transistors, turnstiles and pumps
are explained and the fundamental limits of these devices are discussed in detail.
We describe the various physical mechanisms that influence the device uncertainty
and review the analytical and numerical methods needed to calculate the intrinsic
uncertainty and to optimise the fabrication and operation parameters.
Recent experimental results are evaluated and compared with theoretical predictions.
Although there are discrepancies between theory and experiments,
the intrinsic uncertainty is already small enough to start preparing for
the first SET-based metrological applications.
\end{quote}
\vskip6cm
\begin{center}
DFM-98-R27
\end{center}

\newpage \tableofcontents\newpage

\section{Introduction}

\subsection{General introduction}

In recent years, the development of sub-micron devices has offered the
opportunity to transport electric charge by the manipulation of individual electrons.
Devices have been fabricated that can be controlled by an
externally applied alternating signal in such a way, that an integer number of electron
charges is transferred through the device per cycle. Once the frequency of the external
signal is accurately known, the current through the device can, in principle, be calculated
with high precision. Such single-electron transport (SET) devices have opened the path
towards the realization of high-precision current and capacitance standard.
A high-precision current standard is needed because a standard for the unit of
electrical current, the ampere, is
presently not available in a direct and accurate manner, although it is one
of the base units of the International System of Units (SI).
The construction of a high-precision
capacitance standard using SET devices offers a realistic alternative for
the calculable capacitor. In general, there is also a growing need in
science and industry to measure very low current levels with high precision,
driven by the ongoing tendency towards miniaturization of
electrical circuits.

The basis of high-precision current and capacitance standards that are
presently being developed, is formed by sub-micron devices like turnstiles and
pumps. In practice the observed intrinsic uncertainty of these devices is of
the order of $1$ part in $10^8$, already approaching the uncertainty levels of
Josephson voltage standards and quantum Hall resistance standards.
Theoretically, SET devices are capable to individually transfer electrons
with an error per electron that is much smaller. This discrepancy is caused
by yet unknown mechanisms.

Because of the extremely difficult and failure-sensitive fabrication process
of SET devices and because of the ultra-low temperatures needed for their
operation, research on SET devices is inherently time and money consuming.
An efficient feedback between experimental results and theoretical
understanding is necessary to improve the effectiveness of the research
activities. For that reason, an overview is desired of today's understanding
of the properties of SET devices when applied within metrology.

The focus will be on SET transistors, and
turnstile and pump devices. Their working principle will be discussed in the
first section of this article. A brief introduction to the fabrication
technique and a short overview of the experimental status are given in
Sec. \ref{sec:exp}. In Sec. \ref{sec:errors} and \ref{sec:simulations}
it will be shown that the error mechanisms affecting
the accuray of SET devices are in general reasonably well understood and
that theoretical calculations can support the design of new device lay-outs.
Nevertheless, in Sec. \ref{sec:compare} attention is paid to the fact that in many cases
the experimentally observed uncertainties are still much higher than
expected from theory. Finally, in Sec. \ref{sec:alternatives} some alternative devices are
briefly discussed, such as the recently developed devices that make use of
surface acoustic waves.

Nowadays SET devices are able to accurately generate currents at the level
of 1 pA. This is already high enough to realize a capacitance standard. Such
standard is based on charging a capacitor of typically 1 pF with an
accurately known number of electrons and subsequently measuring the voltage
that has developed across the capacitor. However, the 1 pA level is still
too low to accurately calibrate current sources which operate at higher,
more convenient levels (microamperes and higher). One of the major tasks within SET
metrology is to increase the amplitude of the current that can be generated by
SET devices, while maintaining a very low uncertainty.
This article is meant to serve as a useful reference in that process.

\subsection{Historic framework}

The phenomenon of the Coulomb blockade of tunneling provides the possibility
to control the transport of individual electrons. The Coulomb blockade can
be understood from Fig.\ \ref{fig:settrans} which shows two conducting
electrodes coupled to a small island via tunnel junctions. Transport of
electrons from one electrode to the other can take place by means of
tunneling through the junctions. However, when the total capacitance
$C_{\Sigma }$ of the island is sufficiently small, the energy
$E_{C}=e^{2}/2C_{\Sigma }$ (with $e$ the electron charge) needed to charge
the island with an extra electron becomes high enough to prohibit the
tunneling of other electrons. The Coulomb blockade of tunneling can only be
observed at low temperatures $k_{{\rm B}}T\ll E_{C}$, at which thermally
activated transport is suppressed (here $k_{{\rm B}}$ is Boltzmann's
constant).

The Coulomb blockade was for the first time experimentally observed in
disordered granular materials \cite{zell69}. Later, techniques were
developed to fabricate --- in a controlled way --- ultra-small tunnel
junctions that exhibit the Coulomb blockade effect \cite{fult87}. Nowadays
these junctions form the basis of so-called single electron tunneling (SET)
devices, in which the transport of individual electrons can be controlled by
means of externally applied signals. Likharev {\it et al.} pointed out \cite{likh:rev}
the possibility to use these devices for metrological
applications such as an electrical current standard, for which the magnitude
of the generated current is determined by the frequency $f_{{\rm SET}}$ of
the externally applied signal and the number $k$ of electrons that is
transferred per period:
\begin{equation}
I_{{\rm SET}}=kef_{{\rm SET}}.  \label{Ief}
\end{equation}

The initial suggestion was to pass a current through a single tunnel
junction and to lock the rate $I_{{\rm SET}}/e$ to an external ac source.
The experiments however showed that the Coulomb blockade and consequently
the sought SET oscillations for a single junction are strongly suppressed
\cite{dels89}. This is attributed to quantum charge fluctuations in the
electromagnetic surroundings \cite{naza89,devo90,girv90,flen92girv}.

Special precautions have been taken in several experiments to suppress
quantum fluctuations by attaching highly resistive leads to the junctions
\cite{havi94,kuzm96,zhen98}. Effective shielding from external quantum
fluctuations can also be achieved by placing several tunnel junctions in
series. In this way SET oscillations were experimentally observed in the
form of current plateaux in the current-voltage characteristics
of tunnel junction arrays irradiated by a microwave signal \cite{dels89b}.
However, the flatness of the plateaux was still poor.

Devices to manipulate electrons in a more controlled way were developed,
starting with the so-called single electron turnstile \cite{geer90}. A
slightly more complicated device is the single-electron pump, in which
electrons are moved through a series of tunnel junctions without applying a
bias voltage \cite{poth92}. The first high-accuracy pump was fabricated at
the National Institute of Standards and Technology (NIST) using a 5-junction
design \cite{mart94}. Recently, an improved 7-junction device was fabricated
and studied, and an error per pumped electron of 1 part in 10$^{8}$ was
achieved \cite{kell96,kell97,kell98}. This accuracy is still much lower than
theory predicts on the basis of the present understanding of error
mechanisms. Further research is therefore still necessary. Nevertheless, the
achieved accuracy already corresponds to the required level for the
construction of a new high accuracy capacitance standard
proposed in Ref. \cite{will92}.

Superconducting devices have also been proposed as a candidate to relate
current to frequency \cite{likh85}. Frequency locking at $I=2ef$ has been
observed in small superconducting Josephson junctions (see \cite
{kuzm91,kuzm92} and references therein), but the effect was strongly
suppressed by thermal smearing and quantum fluctuations \cite{naza92,kuzm94}
. Like in the normal metal case, it is necessary to consider arrays of
junctions in order to eliminate the quantum fluctuations \cite{havi96}.

Very recently, a device was developed at Cambridge University that utilizes
a surface acoustic wave (SAW) to pass electrons through a quantum point
contact in the pinch-off regime \cite{shil96b,taly97}. The electric field
induced by the SAW leads to the formation of quantum dots which move along
the constriction transporting an integer number of electrons per period of
the SAW. The first results are encouraging, but much improvement is needed
to achieve an uncertainty comparable to that of a SET pump.

\subsection{Metrological motivation for SET research}

\label{sec:metro}
\begin{figure}[tbp]
\setlength{\unitlength}{0.75cm}
\begin{picture}(10,7)(-5,-1.5)
\put(0,0){\circle{2}}\put(5,5){\circle{2}}\put(10,0){\circle{2}}\put(1,0){\line(1,0){8}}
\put(0.7071,0.7071){\line(1,1){3.5511}}\put(5.7071,4.2929){\line(1,-1){3.5511}}
\put(-0.25,-0.25){\makebox(1,1)[lb]{{\Large \it f}}}
\put(9.75,-.25){\makebox(1,1)[lb]{\Large \it I}}
\put(4.77,4.75){\makebox(1,1)[lb]{\Large\it  V}}
\put(8,2.5){\makebox(1,1)[lb]{\Large $I = i V/R_K$}}
\put(4,-1){\makebox(1,1)[lb]{\Large $I = k e f$}}
\put(-2,2.5){\makebox(1,1)[lb]{\Large $V = nf/K_J$}}
\end{picture}
\caption{The metrological triangle connecting frequency, voltage and current
through the Josephson effect, the quantum-Hall effect and the
single-electron tunneling effect. Here $i,n$ and $k$ are integers and
$R_{K}=h/e^{2}$ and $K_{J}=h/2e$.}
\label{fig:triangle}
\end{figure}
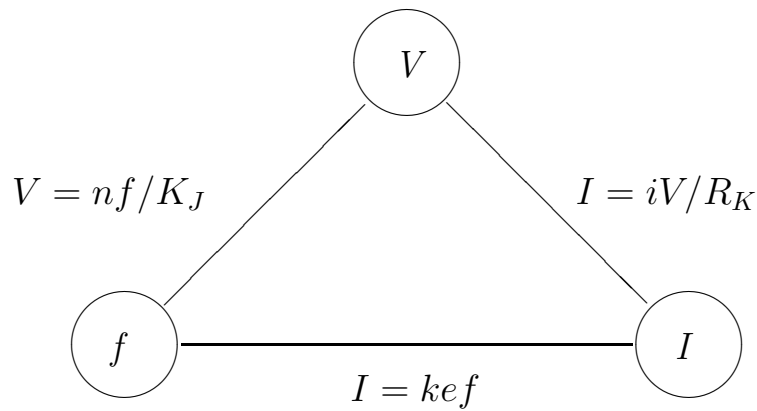

The metrological motivation for research on SET devices lies in their
expected applicability in current standards and capacitance standards.

A current standard based on the manipulation of single electrons will be the
first direct and accurate representation of the ampere. Presently the
ampere, being the (only) electrical base unit within the SI,
is represented by the combination of the
Josephson-array voltage standard and quantized Hall resistance standard
through Ohm's law. This is an indirect and cumbersome method and in practice
only secondary standards for voltage and resistance are combined in this way.

A SET current standard will not be a {\it realisation} of the ampere in the
sense that it is a reconstruction according to the present SI definition,
because the basis for this definition is Ampere's law, which expresses the
force generated by two currents. Ampere's law thus connects the electrical
and mechanical units within the SI system. Using force measurements to
realise the ampere is, however, not very practical. A far more convenient
way appeared to be the application of Ohm's law and to combine voltage and
resistance standards. In 1990 it was agreed that the quantum-Hall effect and
the Josephson effect are representations of the volt and ohm, respectively.
The assignments of values of the new quantum standards were done by
experiments where, e.g., the Josephson voltage was measured in terms of
mechanical units. The resulting relative uncertainties are of the order
$10^{-7}$, which is much larger than the reproducibility of the Josephson
voltage standards, which is of the order $10^{-9}$. The search towards a SET
current standard can therefore also be viewed in a broader perspective,
namely to bring up arguments for a redefinition of the SI system, where the
electrical units are {\it defined} in terms of quantum standards, and
mechanical units are being {\it derived} from those.

Presently, it is possible to make an accurate SET current standard only for
small current values, typically a few picoampere. The uncertainty of the
current is dominated by the uncertainty in the average number $k$ of
electrons that is transferred per cycle of the drive frequency $f_{{\rm SET}}$.
For a SET standard that can be accurately linked to secondary
standards, the current level has to be increased by at least several orders
of magnitude. However, simply increasing the current by means of increasing $
f_{{\rm SET}}$ is not a solution since this would rapidly increase the
relative uncertainty (see Sec. \ref{sec:errors}). It is an inherent drawback
of nowadays current standards that are based on SET devices.

A capacitance standard based on SET devices does not need currents higher
than a few picoampere, in contrast to a current standard. For an application
in capacitance standards, the SET device is used to charge a capacitor with
an accurately known number of electrons \cite{zimm97,will92}. By measuring
the voltage that consequently develops accross the capacitor, the
zero-frequency capacitance value can be determined. Again, the uncertainty
of measurement is dominated by the uncertainty in the average number $k$ of
electrons that is transferred per cycle of the drive frequency $f_{{\rm J}}$.
It is expected that capacitance measurements can be carried out with
relative uncertainties of 10$^{-7}$ at 1 pF \cite{kell:private}.

Besides their potential usefulness as measurement standards for calibration
of current and capacitance, SET devices can also be exploited for
fundamental metrology experiments, such as an alternative determination of
the value of the electron charge. One way of doing this is to use Ohm's law
by passing a SET current through a quantized Hall device while comparing the
Hall voltage with a Josephson-array voltage standard. This closes the
so-called quantum metrological triangle, Fig.\ \ref{fig:triangle}, first
suggested by Likharev and Zorin \cite{likh85}.

The quantized Hall resistance $R$ and the Josephson voltage $V$ are given by
\begin{equation}
V=nf_{{\rm J}}/K_{{\rm J}},\quad R=R_{{\rm K}}/i,  \label{KR}
\end{equation}
where $n$ and $i$ are integers, and $f_{{\rm J}}$ is the drive frequency of
the Josephson-array voltage standard. The constants of proportionality
$K_{{\rm J}}$ and $R_{{\rm K}}$ are called the Josephson constant and the Von
Klitzing constant, respectively. Practical values for $K_{{\rm J}}$ and
$R_{{\rm K}}$ have been recommended to facilitate the international comparison
of voltage and resistance measurements\cite{tayl89}. The recommended
values are denoted by $K_{{\rm J-90}}$ and $R_{{\rm K-90}}$, respectively.
By combining Eqs.~(\ref{Ief}) and (\ref{KR}) and inserting values for
$K_{{\rm J}}$ and $R_{{\rm K}}$ , a value for the electron charge can be
calculated from
\begin{equation}
e=\left( \frac{f_{{\rm J}}}{f_{{\rm SET}}}\right) \times \left( \frac{nk}{i}
\right) \times \left( \frac{1}{R_{{\rm K}}K_{{\rm J}}}\right) .  \label{e90}
\end{equation}
Of course, this value has to be consistent with earlier determinations of $e$.
Based on the physical understanding of the Josephson effect and the
quantum Hall effect, the constants $K_{{\rm J}}$ and $R_{{\rm K}}$ are
predicted to be combinations of $e$ and $h$,
\begin{equation}
K_{{\rm J}}=\frac{2e}{h},\quad R_{{\rm K}}=\frac{h}{e^{2}}\text{.}
\label{KandR}
\end{equation}
This means that, in addition, the value for $f_{{\rm J}}/f_{{\rm SET}}\times
nk/i$ is expected to be exactly equal to 2. It should be stressed that a
comparison of the so measured value of $e$ with the already known value of $e$
does not yield any additional information about the values of the
fundamental constants, or the correctness of the underlying theory. It
merely provides a usefull consistency check, unfortunately without providing
a clue with respect to the source of the possible inconsistencies.

In a more direct way, the electron charge can be measured by charging a
capacitor $C_{s}$ with an accurately known number $m$ of electrons, using a
SET current source. When the value of the capacitor is accurately known,
through traceability towards the calculable capacitor, and by measuring the
voltage across the capacitor using a Josephson standard, the value of $e$
can be calculated from
\begin{equation}
e=\left( \frac{n}{m}\right) \times \left( \frac{C_{s}f_{{\rm J}}}{K_{{\rm J}}}\right)
\text{.}  \label{ecap}
\end{equation}
Again, this value must be consistent with the generally accepted value for
the electron charge. Alternatively this method of charging a capacitor can
be used for a determination of the fine structure constant $\alpha$ \cite{will92},
which is defined by
\begin{equation}
\alpha \equiv \frac{\mu _{0}\,\,c\,e^{2}}{h}\,,  \label{alfadef}
\end{equation}
where $\mu _{0}$ is the permeability of vacuum and $c$ is the velocity of
light. Combining Eqs. (\ref{ecap}) and (\ref{alfadef}) the value for $\alpha$
is given by
\begin{equation}
\alpha =\left( \frac{\mu _{0}\,\,c}{2}\right) \times \left( \frac{n}{2m} \right) \times f_{{\rm
J}}\,C_{s}\text{.}  \label{alfa}
\end{equation}
Such a determination of $\alpha $ would be helpful in the field of
fundamental constants.

\subsection{Basic requirement for the Coulomb blockade of tunneling}

\label{conditions}
The Coulomb blockade of tunneling can only be observed if
the number of electrons on each island is well defined. First of all, this
implies that the energy needed to place an additional charge on an island
must be large compared to the energy related to thermal fluctuations:
\begin{equation}
E_{C}=e^{2}/2C_{\Sigma }\gg k_{{\rm B}}T,  \label{cond1}
\end{equation}
where $C_{\Sigma }$ is the total capacitance of the island. Secondly,
quantum fluctuations must be suppressed. In other words, the wave function
of an electron should be localized on a single island. Therefore, the energy
uncertainty associated with the lifetime $R_{T}C$ due to tunneling of
electrons across the junction (with capacitance $C$ and tunneling resistance
$R_{T}$) must be much smaller than $E_{C}$. From this follows:
\begin{equation}
R_{T}\gg h/e^{2}\simeq 25.8\;{\rm k}\Omega ,  \label{cond2}
\end{equation}
where $h$ is Planck's constant.

The latter condition suggests that the larger the tunneling resistance the
better. This is unfortunately not the case, because the speed of operation
is limited by the time $R_T C$, which therefore should be as short as
possible. One thus has to find an optimum choice of $R_T$ which compromises
between speed and leakage.

\subsection{The SET transistor}

\label{sec:SETtrans}
\begin{figure}[tbp]
\setlength{\unitlength}{0.75cm}
\begin{picture}(15,10)(-5,-1.5)
\put(0,0){\line(1,0){10}}
\multiput(0,0)(5,0){3}{\line(0,1){1}}
\multiput(0,2)(5,0){3}{\circle{2}}
\multiput(0,3)(10,0){2}{\line(0,1){3.5}}
\put(5,3){\line(0,1){1}}
%capacitor plates
\multiput(4,4)(0,0.5){2}{\line(1,0){2}}
\put(5,4.5){\line(0,1){2}}
\put(0,6.5){\line(1,0){2}}\put(3,6.5){\line(1,0){4}}\put(8,6.5){\line(1,0){2}}
%tunnel junctions
\multiput(0,0)(5,0){2}{
\multiput(2,5.5)(0.5,0){3}{\line(0,1){2}}
\multiput(2,5.5)(0,2){2}{\line(1,0){1}}
}
\put(-0.5,1.5){\makebox(1,1)[c]{{\Large $V_L$}}}
\put(4.5,1.5){\makebox(1,1)[c]{{\Large $V_G$}}}
\put(9.5,1.5){\makebox(1,1)[c]{{\Large $V_R$}}}
\put(2.5,3.75){\makebox(1,1)[lb]{{\Large $C_g$}}}
\put(2.25,8){\makebox(1,1)[lb]{{\Large $C_L$}}}
\put(7.25,8){\makebox(1,1)[lb]{{\Large $C_R$}}}
\put(4.5,6.75){\makebox(1,1)[c]{{\Large $Q=ne$}}}
\end{picture}
\vbox to 7cm {\hbox to 5cm {
\includegraphics{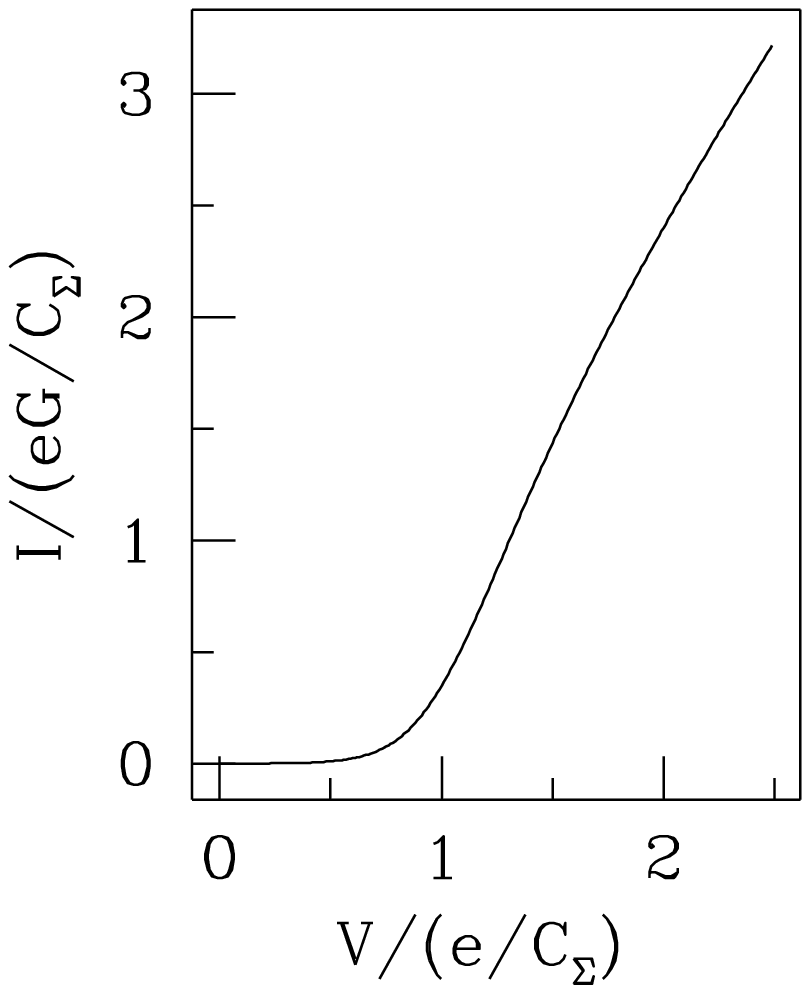}
\includegraphics{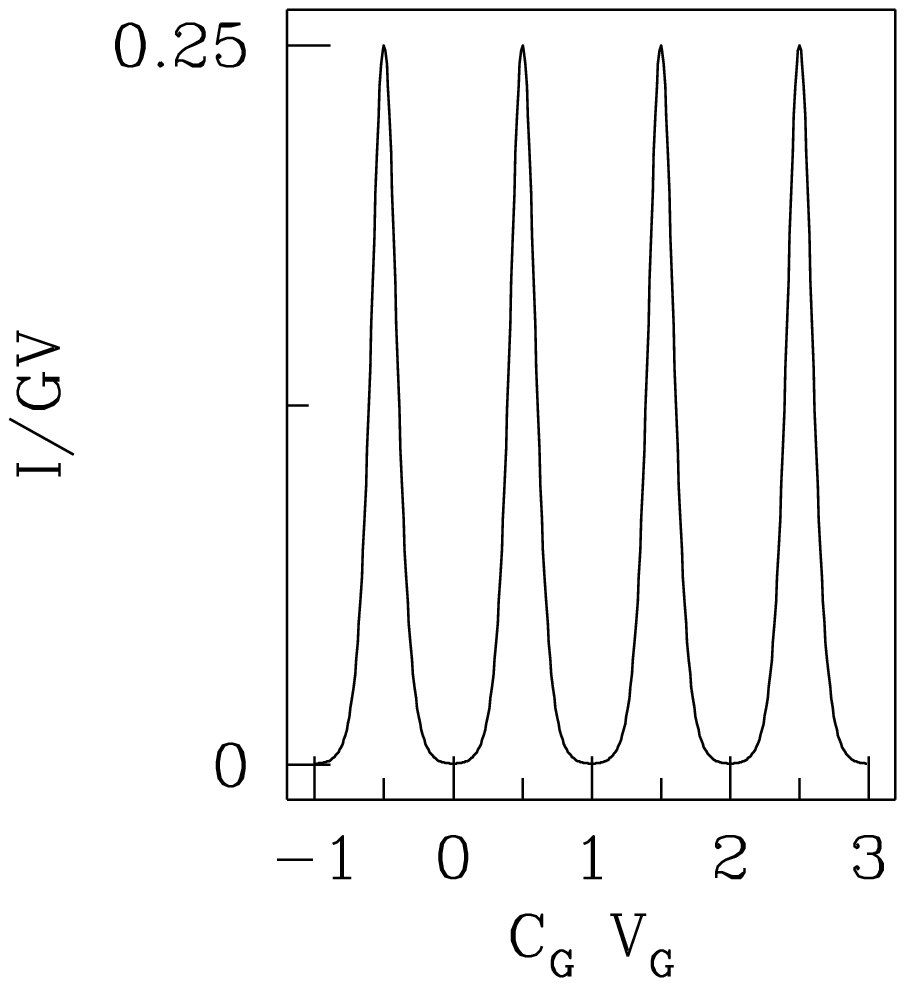}

}}
\caption{SET transistor. The top panel shows a
diagram of a SET transistor with a metalic island which is connected to the
external electrodes through two tunnel junctions (represented by boxes).
The capacitances of the tunnel junctions are $C_{L}$ and $C_{R}$. The lower
left panel shows the calculated current-voltage characteristic at low temperature
($T=E_{C}/10$) and $V_{G}=0$. For a bias voltage above threshold, $V_{tr} =
e/C_{\Sigma}$, the current can flow, whereas for smaller bias voltage, the
current is blocked. The number of electrons on the island can be tuned by
the gate voltage $V_{G}$. The properties of the transistor change
periodically with the gate voltage $V_{G}$. The period corresponds to
addition of one electron to the island. This is shown in the lower right
panel, where the conductance is plotted versus $V_{G}$. The current
reaches a maximum each time the charge state with $N$ electrons has the same energy as
the state with $N+1$ electrons.}
\label{fig:settrans}
\end{figure}

\subsubsection{Properties of the SET transistor}

To elucidate the effect of the Coulomb blockade on the electric properties
of nanometer scale devices we consider the so-called single electron
transistor whose diagram is shown in the upper part of Fig.\ \ref
{fig:settrans}. The central island can only be occupied by an integer number
of electrons, provided that the conditions (\ref{cond1}), (\ref{cond2}) are
fulfilled. By applying a bias voltage ($V_{L}-V_{R}$) and changing the
potential $V_{G}$ of the gate electrode the number of electrons on the
island can be changed. If $V_{G}$ is adjusted such that the charging energy,
needed to add (remove) an electron to (from) the central island, is larger
than the applied bias voltage and the thermal energy, current through the
structure is blocked. If however the potentials are selected so that the
charging energy is zero, current can flow. The device thus acts like a
transistor where a small change in offset charge on the gate capacitor
changes the state of the device from closed to opened. For comprehensive
reviews of the Coulomb blockade see e.g. Refs. \cite{likh:rev,grabert}.

The electrostatics of the system is determined by the external voltages and
the charge, $Q=Ne$, on the central island. The part of the total
electrostatic energy which depends on the number of electrons is
\begin{equation}
E_{\text{el}}=\frac{(Q-Q_{\text{ex}})^{2}}{2C_{\Sigma }},
\label{EelSETtrans}
\end{equation}
where the total capacitance of the island is $C_{\Sigma
}=(C_{L}+C_{R}+C_{G}) $ and the induced charge on the island is defined as
$Q_{\text{ex}}=\left( V_{L}C_{L}+V_{R}C_{R}+V_{G}C_{G}\right) .$ The charging
energies for adding ($\Delta ^{+}$) or removing ($\Delta ^{-}$) an electron
from the island are given by
\begin{equation}
\Delta ^{\pm }=E_{\text{el}}(Q)-E_{\text{el}}(Q\pm e)=\mp \frac{e}{C_{\Sigma
}}\left( Q-Q_{ex}\pm \frac{e}{2}\right) .  \label{Epm}
\end{equation}
Thus if $Q_{\text{ex}}=(n+\frac{1}{2})e$ the two charge states $Q=ne$ and
$Q=(n+1)e$ have the same energy. As a result, current can flow freely because
the system can alternate between the two degenerate states. Away from this
degeneracy point the current is blocked as long as $e(V_{L}-V_{R})\ll
e^{2}/2C_{\Sigma }$ and $k_{{\rm B}}T\ll e^{2}/2C_{\Sigma }$.

The junctions in a SET transistor normally consist of two layers of metal
(usually Al) separated by thin insulating barrier (Al oxide). The overlap of
the initial and final electron states $|i\rangle $ and $|f\rangle $ on the
two sides of the barrier is described by the tunneling Hamiltonian,

\begin{equation}
\hat{V}=\sum_{i,f}T_{f,i}|f\rangle \langle \ i|\;,  \label{Vtunneling}
\end{equation}
where $T_{f,i}$ is the transition probability from state $|i\rangle $ to
state $|f\rangle $. Fermi's golden rule allows one to evaluate the rate
$\Gamma _{if}$ of tunneling from state $|i\rangle $ to state $|f\rangle $,
\begin{equation}
\Gamma _{if}=\frac{2\pi }{\hbar }\left| \left\langle f\left| \hat{V}\right|
i\right\rangle \right| ^{2}\delta \left( E_{f}-E_{i}\right) \;,
\label{GoldenRule}
\end{equation}
Using these formulae, we can find the electron tunneling rates through the
left and right junctions in Fig. \ref{fig:settrans},
\begin{equation}
\Gamma _{L,R}\left( Q\rightarrow Q\pm e\right) =\frac{1}{e^{2}R_{T(L,R)}}
\frac{\Delta ^{\pm }}{1-\exp \left( -\Delta ^{\pm }/k_{B}T\right) }
\label{Gamma}
\end{equation}
where $R_{T(L,R)}=\hbar /2\pi e^{2}|T_{L,R}|^{2}N_{F}^{2}$ are the nominal
tunneling resistances of the junctions and $N_{F}$ is the density of states
at the Fermi level. From Eq.\ (\ref{Gamma}) we see that tunneling is
exponentially suppressed if the electrostatic energy after the tunnel event
is larger than before (i.e. $\Delta ^{\pm }<0$).

It is important to realize that the tunneling rate, given by Eq. (\ref{Gamma}),
relies on the assumption that electrostatic equilibrium is restored
instantaneously after a tunneling event. In terms of time scales this is
true if the characteristic time for redistribution of the charge, which is
the inverse plasma frequency, is much shorter than the ``tunneling time''
$\hbar /\Delta .$ This assumption is sometimes called the capacitance model
(which corresponds to the ``global rule'' in Ref. \cite{grabert}) and it is
certainly a valid approximation for good metal conductors.

The dynamic properties of a SET transistor are determined by the tunneling
rates (\ref{Gamma}), and can be described by the master equation
\begin{equation}
\frac{d}{dt}P(Q,t)=\sum_{\pm }\left[ P(Q\pm e,t)\Gamma (Q\pm e\rightarrow
Q)-P(Q,t)\Gamma (Q\rightarrow Q\pm e)\right] ,  \label{mastereq_a}
\end{equation}
where $\Gamma =\Gamma _{L}+\Gamma _{R}$ is the sum of the tunneling rates
through both junctions and $P(Q)$ is the probability of the charge state $Q$.
In a steady state ($dP(Q,t)/dt=0$) there is a balance between the number
of electrons entering and leaving the island so that the probability of a
transition from the state $Q$ to the state $Q+e$ is equal to the probability
of a transition in the opposite direction. Thus it follows that
\begin{equation}
P(Q)\Gamma (Q\rightarrow Q+e)=P(Q+e)\Gamma (Q+e\rightarrow Q)\;.
\label{balance}
\end{equation}
This equation yields $P(Q)$ and once that is found the current is given by
\begin{equation}
I=e\sum_{Q}P(Q)\left[ \Gamma _{L}(Q\rightarrow Q+e)-\Gamma _{L}(Q\rightarrow
Q-e)\right] .
\end{equation}
In the lower part of Fig.\ \ref{fig:settrans} we illustrate the behavior of
the single electron transistor by plotting the characteristics based on the
solution of Eq. (\ref{balance}). The source-drain conductance depends
strongly on the additional charge on the island induced by the gate voltage.
The distance between two conductance minima corresponds to the addition of
one electron charge to the island. Clearly, a SET transistor is a very
sensitive detector of electric charge.

\subsubsection{Applications of the SET transistor}

\label{sec:appl}Since the source-drain conductance of a SET transistor is
strongly dependent on the induced charge on the island it is possible to use
the SET transistor as a very sensitive detector of changes in gate voltage
\cite{lafa91,fult91}. This can be done by applying a suitable bias current
and by measuring the dependence of the source-drain voltage $V_{SD}$ as a
function of the gate voltage $V_{G}$. After optimisation of the values of
the junction capacitances and gate capacitance a voltage gain $dV_{SD}/dV_{G}$
close to one can be obtained. This electrometer operation of the
SET transistor however only works at low frequencies due to inevitable stray
capacitances. The low-frequency properties are strongly influenced by noise
sources inside and close to the tunnel junctions. These sources give rise to
a $1/f$ - like background charge noise on the gate electrode with a
magnitude of typically $10^{-4}$~$e/\sqrt{{\rm Hz}}$ at 10 Hz. This will be
discussed in Sec. \ref{sec:background}.

The advantage of a SET electrometer over conventional electrometers is their
very low leakage. If the role of stray capacitances can be minimized, SET
electrometers are especially useful when applied as a null detector in the
experiment of charging a cryogenic capacitor using a SET electron pump \cite
{will92}. This was first shown theoretically in Ref. \cite{clar95} and has
recently been confirmed experimentally \cite{kell97}. Another application of
a SET transistor is as a non-invasive probe of the local chemical potential
in two-dimensional electron gas systems \cite{fiel96,wei97}.

The electrometer operation of the single-electron transistor however only
works at low frequencies due to the parasitic lead capacitances which limits
speed. For higher operation speed an on-chip amplifier is necessary. However
recently a new method was developed to avoid this drawback.
Schoelkopf {\it et al.} \cite{scho98} have demonstrated a combination of
a SET transistor and a radio frequency resonant circuit, a so-called RF-SET device.
The SET transistor was used to modify
the damping of the resonant circuit (a similar principle is used in RF
SQUIDs). This RF-SET device has a charge sensitivity of the order 10$^{-5}$
$e/\sqrt{\text{Hz}}$ at 1 MHz, which no doubt will open for new areas of
application of single-electron electrometers, see also Sec. \ref{sec:counter}.

Pekola {\it et al}. showed that a SET transistor can be utilized as a
thermometer \cite{peko94,farh97}. The idea is to use the thermal smearing of
the Coulomb gap in the current-voltage characteristics as a measure of
temperature. In contrast to other single-electron devices, such a
thermometer can operate at temperatures somewhat larger than the temperature
corresponding to the Coulomb charging energy. The advantage of the SET based
thermometer is its insensitivity magnetic fields, unlike other cryogenic
thermometers. In fact this thermometer measures directly the ratio of
temperature to voltage.

Recent technological advances in fabrication have resulted in a SET
transistor positioned at the end of a sharp glass tip \cite{yoo97}. Such a
single-electron scanning electrometer(SETSE) is capable of mapping static
electric fields and charges with a spatial resolution of 100 nm and a charge
sensitivity of a small fraction of an electron charge. The SETSE has been
used to image and measure depleted regions, local capacitance, band bending,
and contact potentials on the surface of semiconductor samples. Another application of
a SET transistor is as a non-invasive probe of the local chemical potential
in two-dimensional electron gas systems \cite{fiel96,wei97}.

\subsection{Single-electron pumps and turnstiles}

\label{sec:pumpturnstile}
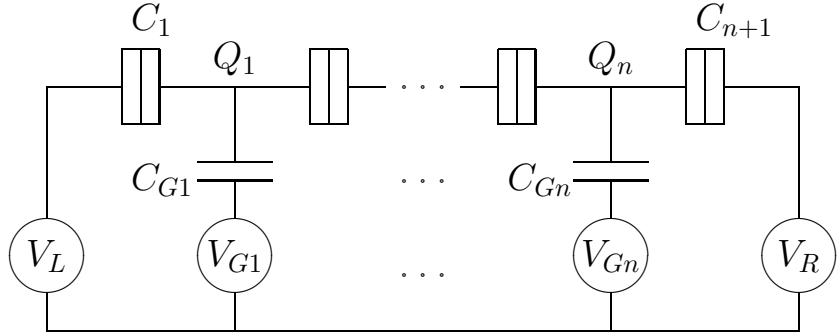
\begin{figure}[tbp]
\setlength{\unitlength}{0.5cm}
\begin{picture}(10,10)(-5,-1.5)
%tunnel junctions
\multiput(0,0)(5,0){4}{
\multiput(2,5.5)(0.5,0){3}{\line(0,1){2}}
\multiput(2,5.5)(0,2){2}{\line(1,0){1}} }
%Voltage sources
\multiput(0,2)(5,0){2}{\circle{2}}\multiput(15,2)(5,0){2}{\circle{2}}
\multiput(0,0)(5,0){2}{\line(0,1){1}}\multiput(15,0)(5,0){2}{\line(0,1){1}}
\put(0,0){\line(1,0){20}}\multiput(5,0)(7,0){2}{\line(1,0){3}}
%capacitor plates and lines
\multiput(0,0)(10,0){2}{
\multiput(4,4)(0,0.5){2}{\line(1,0){2}}\put(5,3){\line(0,1){1}}
\put(5,4.5){\line(0,1){2}}\put(3,6.5){\line(1,0){4}} }
\multiput(0,6.5)(18,0){2}{\line(1,0){2}}\multiput(0,3)(20,0){2}{\line(0,1){3,5}}
\multiput(0,0)(0,2.5){3}{\multiput(9.5,1.5)(0.5,0){3}{\circle{.1}}}
\multiput(8,6.5)(3,0){2}{\line(1,0){1}}
%text
\put(-0.5,1.5){\makebox(1,1)[c]{{\Large $V_L$}}}
\put(4.5,1.5){\makebox(1,1)[c]{{\Large $V_{G1}$}}}
\put(14.5,1.5){\makebox(1,1)[c]{{\Large $V_{Gn}$}}}
\put(19.5,1.5){\makebox(1,1)[c]{{\Large $V_R$}}}
\put(2.25,3.75){\makebox(1,1)[lb]{{\Large $C_{G1}$}}}
\put(12.25,3.75){\makebox(1,1)[lb]{{\Large $C_{Gn}$}}}
\put(2.25,8){\makebox(1,1)[lb]{{\Large $C_1$}}}
\put(17.25,8){\makebox(1,1)[lb]{{\Large $C_{n+1}$}}}
\put(4.5,6.75){\makebox(1,1)[c]{{\Large $Q_1$}}}
\put(14.5,6.75){\makebox(1,1)[c]{{\Large $Q_n$}}}
\end{picture}
\caption{Principle of pump. A series of islands
$i=1,\ldots ,n$, with charges $Q_{i}$, are connected by tunnel junctions and
coupled through capacitances $C_{Gi}$ to gate voltages. These gate voltages
control the charges on the islands. By suitable periodic variations of the
gate voltages a well-defined number of electrons can be transfered through
the device. The turnstile has a finite voltage drop across the array, and
typically only one time-dependent gate voltage is applied to the central
island. The pump usually operates without a dc bias.}
\label{fig:genpump}
\end{figure}
Single-electron pumps and turnstiles are devices that generate a dc current
satisfying the fundamental relation (\ref{Ief}) with high accuracy. They
consist of a series of tunnel junctions in which the electric potential of
the islands are manipulated such that only one electron is transferred
through the device per cycle of an externally applied signal. The general
layout for both devices is shown in Fig.\ \ref{fig:genpump}. The islands
with charges $Q_{1},\ldots ,Q_{n}$ ($Q_{i}=N_{i}e$) are separated by tunnel
junctions with capacitances $C_{1},\ldots ,C_{n+1}$. The electric potential
of each island can be controlled by the gate voltages $V_{G1},\ldots ,V_{Gn}$
through the gate capacitances $C_{G1},\ldots ,C_{Gn}$.

The dynamics of such a device can be described by a master equation for the
probabilities $P({\bf Q},t)$ of the charge states ${\bf Q}=(Q_{1},\ldots
,Q_{n})$ of the system \cite{glaz88matv,aver90koro,been91}, which is a
generalisation of Eq. (\ref{mastereq_a}) for the non-stationary case
\begin{equation}
\frac{d}{dt}P({\bf Q},t)=\sum_{{\bf Q}^{\prime }}\left[ P({\bf Q^{\prime }},t)
\Gamma ({\bf Q^{\prime }}\rightarrow {\bf Q})-P({\bf Q},t)\Gamma ({\bf Q}
\rightarrow {\bf Q^{\prime }})\right] .  \label{masterequation}
\end{equation}
The first term comes from tunneling {\it \ into } the charge state ${\bf Q}$
with a rate $\Gamma ({\bf Q^{\prime }}\rightarrow {\bf Q})$ whereas the
second term is due to tunneling {\it \ out of } this state with a rate
$\Gamma ({\bf Q}\rightarrow {\bf Q}^{\prime })$. The tunneling rates $\Gamma $
are time dependent if $\Delta ^{\pm }$ is time dependent, see Eq. (\ref
{Gamma}). Numerical simulations based on this master equation are reviewed
in Sec. \ref{sec:simulations}.

Notice that we described the single-electron tunneling process in terms of
the orthodox rate equation (\ref{masterequation}), which assumes
thermalization of the system between each tunneling event. If the system does
not have time to equilibrate between tunneling events the non-equilibrium
distribution of the electron system has to taken into account. This has been
considered in Refs. \cite{koro94,liu97}.

\subsubsection{The electron pump}

\label{sec:pump}
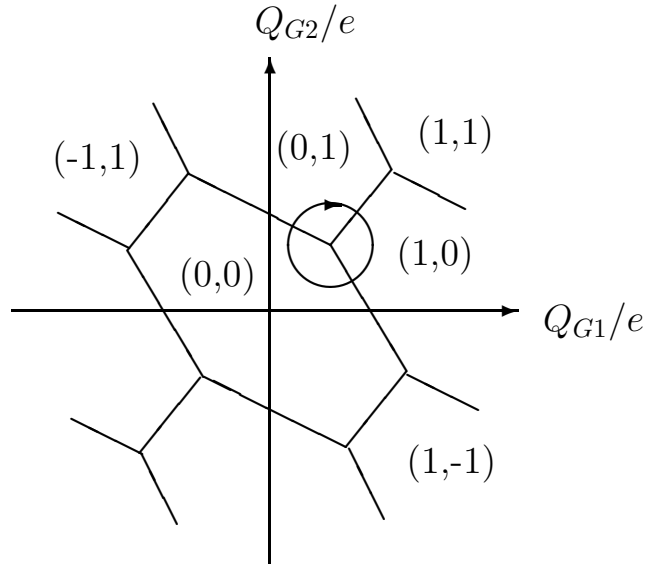
\begin{figure}[tbp]
\setlength{\unitlength}{0.0015cm}
\begin{picture}(5000,5000)(0,-6000)
\thicklines
\put(2251,-3661){\vector( 1, 0){4500}}\put(4546,-5911){\vector( 0, 1){4500}}
\put(5086,-3076){\line( 3,-5){675}}\put(3279,-3125){\line( 3,-5){675}}
\put(5086,-3076){\line(-2, 1){1260}}\put(5221,-4876){\line(-2, 1){1260}}
\put(5761,-4201){\line(-4,-5){540}}\put(3821,-2442){\line(-4,-5){540}}
\put(5626,-2401){\line(-4,-5){540}}\put(3927,-4255){\line(-4,-5){540}}
\put(3826,-2446){\line(-1, 2){315}}\put(3295,-3103){\line(-2, 1){630}}
\put(5086,-3076){\circle{726}}\put(5626,-2401){\line(-1, 2){315}}
\put(5076,-2720){\vector(1,0){100}}
\put(5563,-5515){\line(-1, 2){315}}\put(3718,-5560){\line(-1, 2){315}}
\put(3412,-4939){\line(-2, 1){630}}\put(6283,-2752){\line(-2, 1){630}}
\put(6400,-4543){\line(-2, 1){630}}
\put(5671,-3256){\makebox(0,0)[lb]{\Large (1,0)}}
\put(4591,-2311){\makebox(0,0)[lb]{\Large(0,1)}}
\put(3736,-3436){\makebox(0,0)[lb]{\Large(0,0)}}
\put(5851,-2176){\makebox(0,0)[lb]{\Large(1,1)}}
\put(5761,-5101){\makebox(0,0)[lb]{\Large(1,-1)}}
\put(2611,-2401){\makebox(0,0)[lb]{\Large(-1,1)}}
\put(4411,-1186){\makebox(0,0)[lb]{\Large $Q_{G2}/e$}}
\put(6976,-3841){\makebox(0,0)[lb]{\Large $Q_{G1}/e$}}
\end{picture}
\caption{Configuration space for the simplest electron pump operating with
two gates. A single electron transfer through the device is obtained when
the gate charges follow the circle one turn. }
\label{fig:2pump}
\end{figure}

The simplest electron pump contains two islands ($n=2$) and three tunnel
junctions (see Fig.\ \ref{fig:genpump}). By changing the gate potentials
$V_{G1}$ and $V_{G2}$ the charge configuration ${\bf Q=(}Q_{1},Q_{2})$ can be
controlled.
Pump operation starts by increasing the potential of the island "1" such
that an electron tunnels onto this island from the left electrode. Then the
potential of the island "1" is lowered again, while simultaneously
increasing the potential of the island "2". This results in tunneling of an
electron from the island "1" to the island "2". Finally, by lowering the
potentials of both islands, an electron from the island "2" will be forced
to tunnel to the right electrode. The device has then returned to its
initial charge state and one electron has been pumped from left to right.
Typical cycling rates are in the range of a few megahertz, i.e. RF
frequencies, corresponding to currents of the order of a picoampere. For
optimal performance the bias voltage ($V_{L}-V_{R}$) must be near zero.

One can illustrate the pump operation by means of the diagram in Fig. \ref
{fig:2pump}, which displays the regions of minimum Coulomb energy in the
$(Q_{G1},Q_{G2})$ plane for the different charge states ($Q_{1},Q_{2}$),
where $Q_{Gi}=C_{Gi}V_{Gi}$, ($i=1,2$). The Coulomb energy $E_{\text{el}}$
is given by
\begin{equation}
E_{\text{el}}=\frac{1}{3C}\left[
(Q_{1}-Q_{G1})^{2}+(Q_{2}-Q_{G2})^{2}+(Q_{1}-Q_{G1})(Q_{2}-Q_{G2})\right] \;.
\label{Estatpump}
\end{equation}

If two harmonic signals, shifted in phase by $\pi /2$, are applied to the
gates the system will follow a circular path in the ($Q_{G1}, Q_{G2}$)
plane. The type of pump operation described above corresponds to the
situation where the circle encloses one of the triple points ({\it viz}. the
transfer between the states $(0,0)$, $(1,0)$, and $(0,1)$, as shown in Fig.\
\ref{fig:2pump}). The path in the ($Q_{G1}, Q_{G2}$) plane can be chosen in
many ways and the optimum choice will be discussed in section Sec. \ref
{sec:results}.

\subsubsection{The turnstile}

\label{sec:turnstile}
\begin{figure}[tbp]
\epsfxsize=15cm\epsfbox{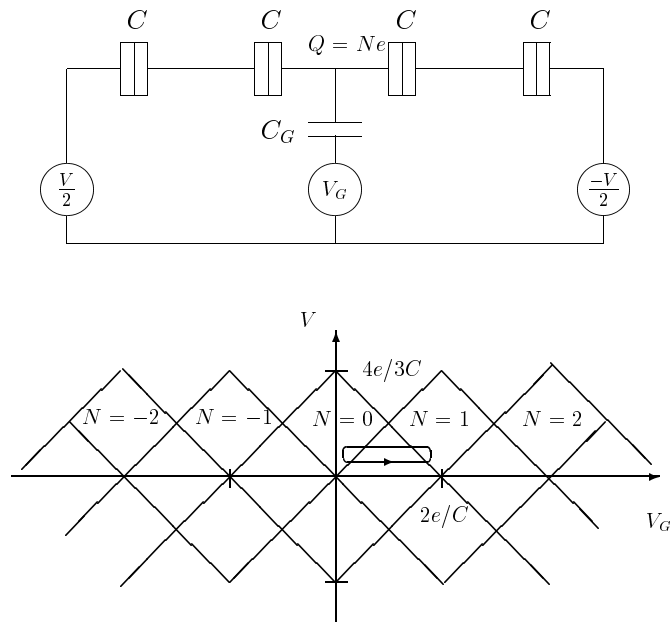} \vspace{-10cm}
\caption{Schematic layout of the single-electron turnstile device. The device has
four small tunnel junctions and a gate capacitance. In the lower part of the
figure, the stability diagram in the $(V_{G}, V)$ plane is shown. Each
square represents a stable configuration with $N$ electrons on the
middle island. The ratio of the tunnel junction capacitance and the gate
capacitance has the optimum value: $C/C_{G}=2$. One turn around the
operation trajectory depicted in the lower part, transfers one electron
through the device in the direction given by the bias voltage. }
\label{fig:turnstile}
\end{figure}

The simplest electron turnstile contains three islands and four tunnel
junctions (upper part of Fig.~\ref{fig:turnstile}). The turnstile differs
from the electron pump by the fact that the external drive frequency needs
to be applied only to one gate electrode coupled to the middle island. On
the other hand, a non-zero bias voltage is required to define the direction
of the current. A single drive signal is an advantage compared to the more
complex pump operation. The finite bias is a disadvantage since it causes
energy dissipation in the device. At the very low temperatures, required for
the proper operation of SET devices, this can be a major problem since the
transfer of heat from the metal islands to the substrate is poor which
causes heating of the electron gas. This makes the turnstile less favourable
for high precision usage.

The operation of the turnstile is illustrated in the lower part of Fig.\ \ref
{fig:turnstile}, which displays regions of minimum Coulomb energy in the
$(V_{G}, V)$ plane for different numbers $N$ of additional electrons on the
middle island. For a finite value of the bias voltage $V=(V_{L}-V_{R})$ and
$V_{G}=0$, the turnstile is in the $N=0$ region. The turnstile cycle starts
by increasing the potential of the gate electrode. When the system leaves
the $N=0$ region, an electron tunnels from the left electrode, via the two
junctions on the left, onto the middle island. This brings the system to the
$N=1$ state. By subsequently lowering the gate potential, the system returns
to the $N=0$ state as an electron tunnels from the middle electrode, via the
two junctions on the right, to the right electrode. In this way a single
electron charge is guided through the turnstile for every cycle of the gate
voltage.

\section{Experimental status of pumps and turnstiles}

\label{sec:exp}

\subsection{Fabrication technique}

\label{sec:fab}
\begin{figure}[tbp]
\epsfxsize=15cm \epsfbox{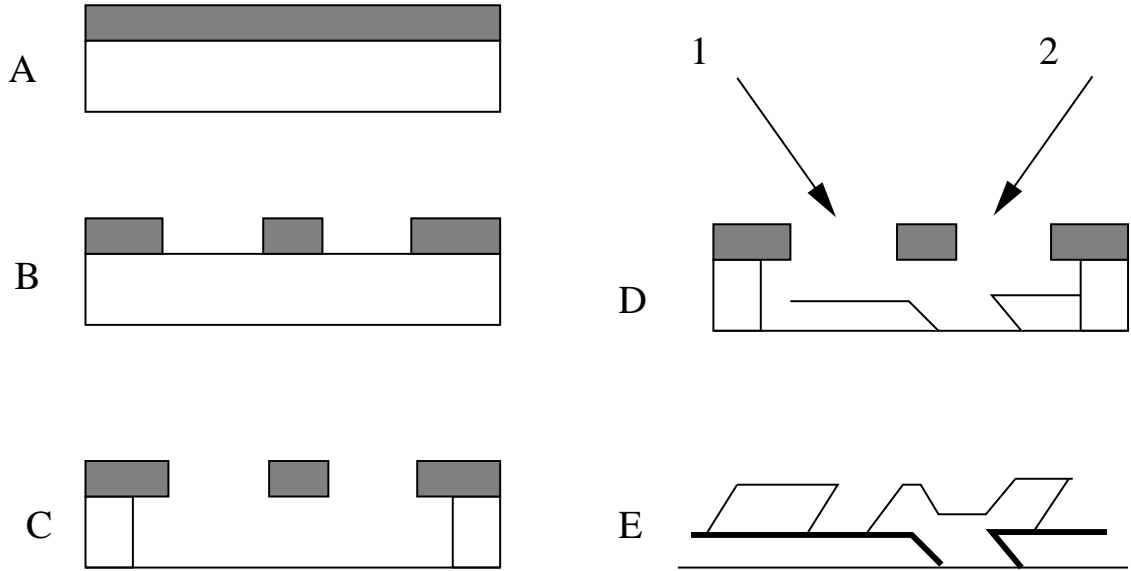}
\caption{Illustration of the double angle evaporation technique. Starting
with a substrate covered with resist (A), a pattern is written using
electron beam lithograpy. After development (B), the structure is etched in
several steps, such that a free-hanging bridge is created (C). A metal layer
is then evaporated (D) from angle 1, and after oxidation and evaporation of
the second metal layer from angle 2, the material used to form the mask is
removed and the tunnel junction is created (E). Note that due to the double
angle evaporation two copies of each feature are produced, as can be seen in Fig.\
\ref{fig:NISTpump}. }
\label{fig:fab}
\end{figure}

In order to actually realise the devices that were discussed in Secs.\ \ref
{sec:SETtrans} and \ref{sec:pumpturnstile}, the conditions set in Eqs. (\ref
{cond1}) and (\ref{cond2}) must be met experimentally. The smallest
structures, with known geometry and well controlled tunneling resistances,
that can be fabricated using nanoscale technology yield capacitances of the
order of parts of a femtofarad. This means that the operating temperatures
must be significantly lower than 1 K, which can be obtained by the use of a
dilution refrigerator.

The commonly used fabrication technique for SET devices is the so-called
double angle evaporation technique. A resist layer is applied onto some
substrate. Using electron beam lithography a pattern is written in the
resist layer. After development of the resist the resulting nanometer
pattern is used as a mask for the evaporation of (usually) Al. The
evaporation is performed under two angles with an intermediate oxidation
step. The tunnel barriers are formed by the oxidization of the first
aluminum layer. The technique is clarified in Fig. \ref{fig:fab}. Nowadays,
tunnel junctions with typical dimensions of 50 $\times $ 50 nm$^{2}$ can be
made more or less routinely. Using this technique, different types of metals
can be applied to make a varyity of junction types: normal-insulating-normal
junctions (NIN, e.g., Al/Al oxide/Al junctions exposed to a small magnetic
field), superconducting-insulating-superconducting junctions (SIS) and
superconducting-normal-superconducting junctions (SNS). Transport through
the latter two types of junctions takes place by means of Cooper pairs.

\begin{figure}[tbp]
%\begin{picture}(10,5)(-5,-1.5)
%\put(0,0){\mbox{\large Atomic Force picture}}
%\end{picture}
\vbox to 15cm {\hbox to 15cm {
\includegraphics{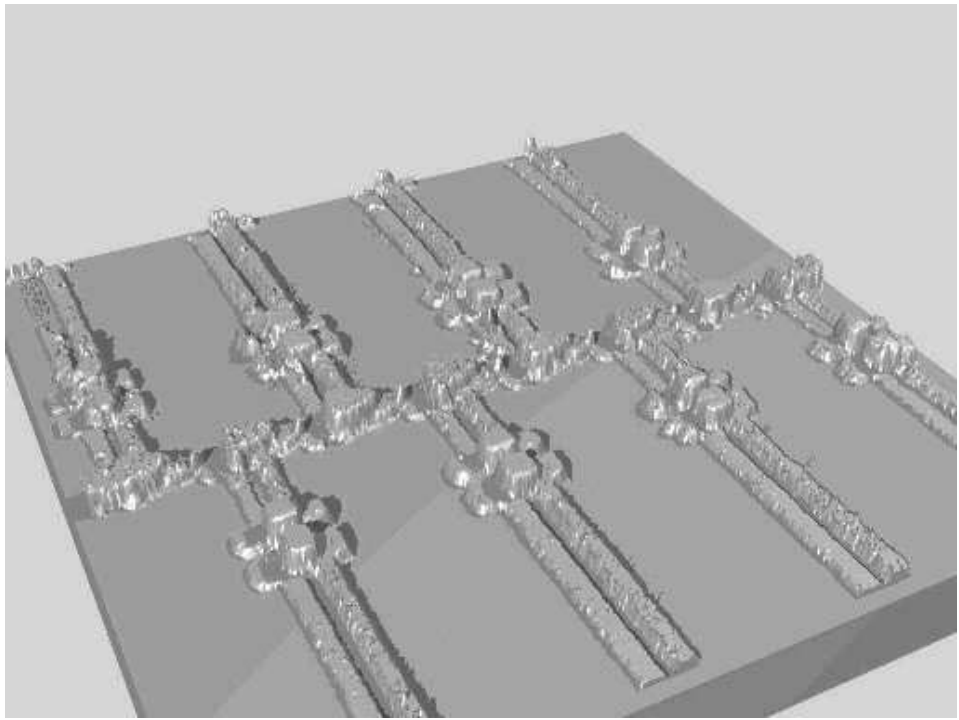} }}
\caption{Atomic force microscop image of a part of a seven junction
single-electron pump \protect\cite{kell98}. The
shadow evaporation explained in Fig.\ \ref{fig:fab}, results in a structure
where each feature appears twice (e.g. the two parallel leads leading to the
capacitances). The size of the island is approximately 700 $\mu$m, and the
heigth of each layer is approximately 50 nm}
\label{fig:NISTpump}
\end{figure}

\subsection{Uncertainty}

\label{sec:NIST}
\begin{figure}[tbp]
\vbox to 10cm {\hbox to 10cm {
\includegraphics{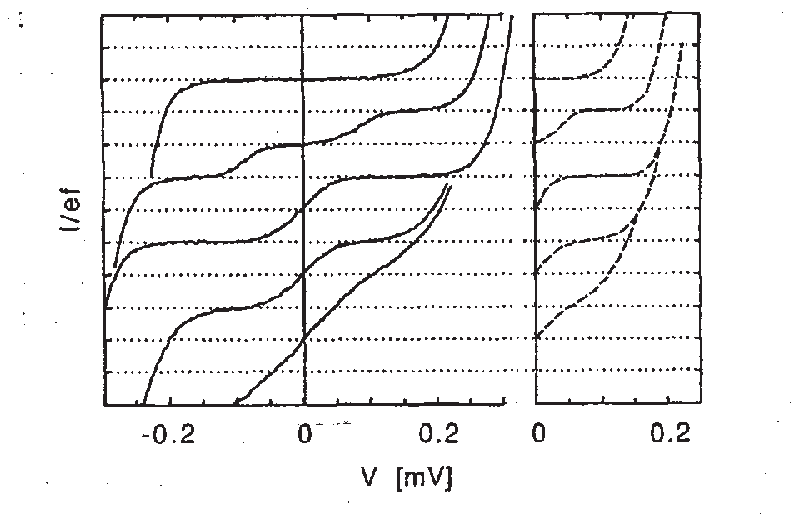} }}
\caption{Current-voltage characteristics for turnstiles by Geerligs {\it et
al.} taken from Ref. \protect\cite{geer90}. The left panel shows
experimental data. The right panel is a comparison with numerical solutions
of the master-equation. The lower curve is without applied ac potential on
the gate and the next ones are for increasing ac power. Dotted lines
indicate the expected plateau value, $I=ef$ for the 5 MHz applied signal. }
\label{fig:expturnstile}
\end{figure}

In the early 90's the first turnstiles were made at Delft University of
Technology \cite{geer90,ande90,kouw91} and the first electron pumps were
fabricated at the Centre d'Etudes Nucl\'{e}aires de Saclay \cite{poth92,urbi91}.

Measurements on a four-junction turnstile \cite{geer90} is shown in the left
part of Fig. \ref{fig:expturnstile}. In the right part of the figure,
calculated $I-V$ curves are plotted. The calculations are based only on
classical tunneling probabilities (see Eq. (\ref{Gamma})), not taking into
account cotunneling (see Sec. \ref{sec:errors}). The value of the generated
current when operating at the plateau agreed with $I=ef$ within the
measurement uncertainty of 0.3 \%. Despite the first promising results, no
determination of the turnstile current with a higher accuracy has been
reported since 1990, even though the group in Delft University of Technology
later continued the work on turnstiles \cite{verbrugh} using more advanced
fabrication techniques.

\begin{figure}[tbp]
\vbox to 10cm {\hbox to 10cm {
\includegraphics{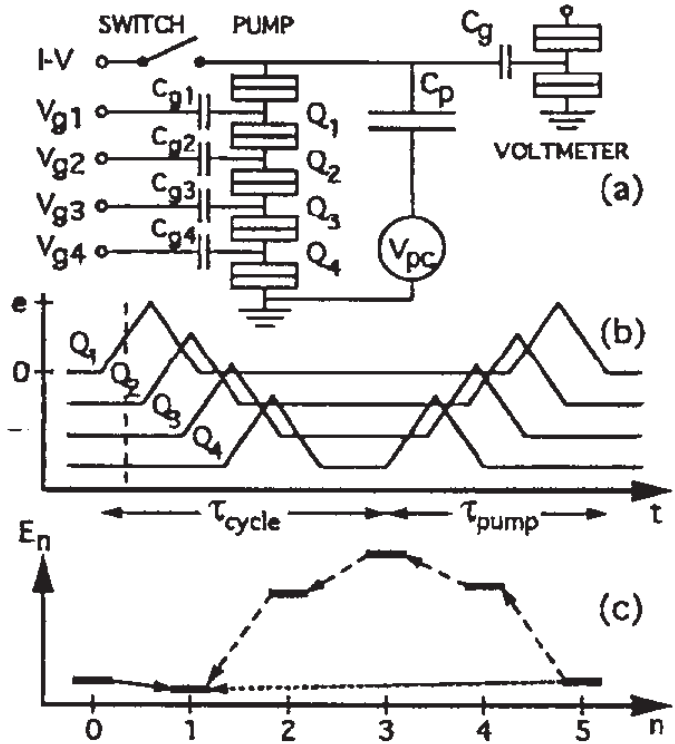} }}
\caption{Electrical diagram and operational principle of a 5-junction
pump, taken from Ref. \protect\cite{mart94}; (a) shows the equivalence
diagram, (b) the sequence of charge polarizations applied to the
islands (the traces are offset for clarity), (c) the Coulomb energy $E_{n}$
versus an electron on island island $n$ at time indicated by dashed line in
(b). The solid arrow is the desired tunneling, while dashed and dotted arrows
show unwanted thermal or cotunneling processes, respectively. }
\label{fig:5pump}
\end{figure}

In 1994 a 5-junction electron pump was fabricated at NIST with an accuracy,
expressed in terms of the error per pumped electron, of 0.5 part in 10$^{6}$
\cite{mart94}. The equivalent diagram of this pump is shown in Fig. \ref
{fig:5pump}(a). The sequence of triangular-shaped voltage pulses, optimized
according to Ref. \cite{jens92}, was applied to the four gate electrodes,
shown in Fig. \ref{fig:5pump}(b). The pulses were carefully designed to
compensate for cross-capacitances. The electron pump was used to transfer an
electron to and from the neighboring island, so that the average charge
transport in time is zero. The charge on the island was monitored by a SET
transitor, which was capacitively coupled to the island. In this way the
error rate of the pump was determined.

The next generation of pumps made at NIST contained 7 junctions \cite
{kell96,kell97,kell98}. In Fig.\ \ref{fig:NISTpump} an atomic force
microscop picture of such a pump is shown. For this type of pump the error
per pumped electron was reduced to 1 part in 10$^{8}$. This intrinsic error
rate is already low enough to try to realize a capacitance standard as discussed in
Sec. \ref{sec:metro}.

\section{Error mechanisms}

\label{sec:errors}There are a number of error mechanisms which deteriorate
the high-accuracy operation of single-electron devices. Some of these errors
are due to classical effects (cycle missing and thermally activated
tunneling), whereas others are due to the quantum mechanical nature of the
electron transport (co-tunneling and photon assisted tunneling), which
becomes increasingly important at low temperatures. In the next subsections
we explain the physics of these error events. We will not review the
calculation of the combined error rates. For those results, we refer to the
original articles, e.g. the table in Ref. \cite{jens92} and the numerical
works reviewed in Sec. \ref{sec:simulations}.

\subsection{Cycle missing}

\label{sec:cycle} When a single electron device is driven by signals of
sufficiently low frequency, the distribution of electrons on the islands
will at any time correspond to the state with the lowest energy. This is
called the {\it adiabatic} regime. At higher frequencies the electrons are
not given enough time to tunnel to the state with lower energy and the
device operates {\it non-adiabatically}. In other words, in this regime a
so-called cycle missing may occur. For example, in the electron pump a
tunneling event may be missed when we circulate around the triple points as
shown in Fig.\ \ref{fig:2pump} and the boundaries between the different
charge states are crossed too rapidly. This means that on the average less
than one electron is transferred per cycle, which leads to deviation from
the ideal behaviour as described by Eq. (\ref{Ief}).

The analysis of a 3-junction electron pump with harmonic drive has been
carried out in Refs.\ \cite{poth91,iwas95}. It was found that reliable
operation is only possible for frequencies much smaller than $1/(R_{T}C)$,
where $C$ is the junction capacitance. For an $N$-junction pump, the error
can be estimated by calculating the probability that an electron does not
tunnel during the time where one of the gate charges, say $Q_{1}$, raises
from 0 to $e$. This results in a error rate per cycle of  \cite{jens92}
\begin{equation}
\varepsilon =\exp \left( -\frac{N-1}{8N^{2}R_{T}Cf}\right) \;,
\end{equation}
where $f$ is the operation frequency. From these expression it is clear that
it is the $R_{T}C$-time that sets the limit for the operation speed. A small
$R_{T}C$ time is desirable for high speed performance, but, on the other
hand, a small tunnel resistance $R_{T}$ will lead to large quantum
fluctuations and unwanted tunneling events such as cotunneling, described in
Sec.\ \ref{sec:cotunneling}.

In order to minimize the cycle missing error it is important to optimize the
waveform of the applied signals. In this way the time spent in a state where
the expected tunneling is energetically favorable, can be maximized. For the
turnstile a square-type wave form rather than a sinusoidal wave form is
preferable. For the pump a triangular signal with optimized amplitudes is
used in metrological experiments \cite{mart94,kell96,kell97,kell98}. Further
optimization of the waveform using numerical algorithms is reviewed in Sec.
\ref{sec:results}.

\subsection{Thermally activated errors}

\label{sec:thermal}

Thermally activated tunneling is the main source of errors at high
temperatures. In Fig.\ \ref{fig:5pump}, a series of thermally enhanced
tunneling events is indicated by the dashed arrow. From Eq.\ (\ref{Gamma})
it follows that the tunneling rate is suppressed exponentially if the energy
of the final state is much larger than the energy of the initial state,
\begin{equation}
\Gamma _{\text{thermal}}\approx \frac{\left| \Delta ^{\pm }\right| }{
e^{2}R_{T}}\exp \left( -\left| \Delta ^{\pm }\right| /k_{B}T\right) \;.
\end{equation}
One can thus calculate the probability of a sequence of tunneling the
events. By taking into account the most important sequences, Jensen and
Martinis \cite{jens92} found approximate formulas for the error rate per
cycle due to thermally activated processes
\begin{equation}
\varepsilon _{\text{thermal}}\approx c\exp \left( -d\frac{E_{c}}{k_{B}T}
\right) \;,
\end{equation}
where the parameters $c$ and $d$ depend on the number of junctions and the
bias voltage. This expression was confirmed by numerical simulations. For
the parameters of the accurate pump described in Sec.\ \ref{sec:NIST}, the
thermal errors dominate for temperatures larger than 100 mK, but are
strongly suppressed for lower temperatures.

If the system does not have to relax between tunnel events the electron gas
heats up and consequently  the electron temperature is higher than the
lattice temperature.  The elevated temperature is a result of a balance between
input power and the relaxation due to electron-phonon coupling\cite{liu97,koro94}.
The cooling power is decreasing at low temperares as a power law approximately
 as $\Sigma VT^5$, where
$V$ is the volume and $\Sigma$ is a dissipation constant. On the other hand, the input
power is proportional to the driving freqeuncy and the charging energy, $fE_C$. Using
this line of reasoning an estimated electron temperature of 80 mK  was found in Ref.\ \cite{liu97}
using typical parameter for metalic single electron pump devices, $f=10$ MHz and a base
temperature of 10 mK. From this consideration it is evident that there is a limit to the lowest
operation temperature for single electron devices.

The understanding of the thermal errors due to tunneling is on firm ground.
However, thermal effects may also be important for processes which involve higher order
tunneling processes such as thermally activated cotunneling. This is
explained in the next section.

\subsection{Cotunneling}

\label{sec:cotunneling}
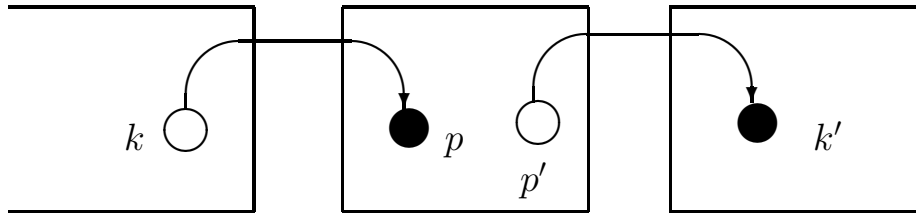
\begin{figure}[tbp]
\setlength{\unitlength}{0.0008in}
\begin{picture}(6029,1394)(1500,-4988)
\thicklines
\put(7201,-4246){\vector( 0,-1){0}}
\put(6481,-4246){\oval(1440,900)[tr]}
\put(6481,-4246){\oval(1440,900)[tl]}
\put(3466,-4426){\circle{284}}
\put(4941,-4416){\circle*{284}}
\put(5790,-4377){\circle{284}}
\put(7238,-4379){\circle*{284}}
\put(4501,-3616){\line( 0,-1){1350}}
\put(4906,-4291){\vector( 0,-1){0}}
\put(4186,-4291){\oval(1440,900)[tr]}
\put(4186,-4291){\oval(1440,900)[tl]}
\put(2296,-3616){\line( 1, 0){1620}}
\put(3916,-3616){\line( 0,-1){1350}}
\put(3916,-4966){\line(-1, 0){1620}}
\put(2296,-4966){\line( 0, 1){0}}
\put(2296,-4966){\line( 0, 1){0}}
\put(4501,-3616){\line( 1, 0){1620}}
\put(6121,-3616){\line( 0,-1){1350}}
\put(6121,-4966){\line(-1, 0){1620}}
\put(4501,-4966){\line( 0, 1){0}}
\put(4501,-4966){\line( 0, 1){0}}
\put(8281,-3616){\line(-1, 0){1620}}
\put(6661,-3616){\line( 0,-1){1350}}
\put(6661,-4966){\line( 1, 0){1620}}
\put(8281,-4966){\line( 0, 1){0}}
\put(8281,-4966){\line( 0, 1){0}}
\put(5671,-4831){\makebox(0,0)[lb]{{\Large $p'$}}}
\put(3061,-4561){\makebox(0,0)[lb]{{\Large $k$}}}
\put(5176,-4561){\makebox(0,0)[lb]{{\Large $p$}}}
\put(7606,-4561){\makebox(0,0)[lb]{{\Large $k'$}}}
\end{picture}
\caption{Cotunneling event in a SET transistor. Two electrons
tunnel in a coherent way such that the charge on the island is the same in
the initial and the final states. Therefore the process overcomes the
Coulomb energy barrier, but is suppressed due to limited number of
available final states in phase space. Furthermore, the probability of the
two electron process is proportional to the product of the two
tunneling conductances. }
\label{fig:cotun}
\end{figure}
In the master-equation described in Sec. \ref{sec:pumpturnstile}, we have
sofar included only the processes where the electron tunneling through a
single junction is involved. However, higher order processes are also
possible \cite{aver89}. One such process is a cotunneling event where two or
more electrons tunnel simultaneously through several junctions in a
quantum-coherent manner.

In the case of the SET transistor, explained in Sec.\ \ref{sec:SETtrans} a
two electron cotunneling process can transfer an electronic charge from left
lead to the right lead such that the number of electrons on the island is
the same in the initial and the final state. This process is illustrated in
Fig.\ \ref{fig:cotun}. There are two possible intermediate states, one where
an electron jumps through the right barrier first leaving an excess hole on
the island and one where an electron jumps through the left barrier first.

There are two fundamentally different types of cotunneling processes: if the
state of the electron that enters the island is identical to the state of
the electron that leaves the island the process is called an {\it elastic}
cotunneling process whereas if the two states are different it is an {\it inelastic}
process. The latter is generally dominant for the metallic
systems where the island has a large number of electronic states \cite{aver90}.

In Section \ref{sec:SETtrans} we derived the tunneling rate using Fermi's
golden rule. The cotunneling rates can be derived in a similar fashion
except that now we must include processes of higher order in the tunneling
matrix elements which connect electronic states across the tunnel junctions.
The effective matrix elements for the inelastic cotunneling process depicted
in Fig. \ref{fig:cotun} can be found using quantum mechanical perturbation
theory. Since it is a quantum mechanical process energy need not be
conserved in the intermediate state, i.e., the state immediately after the
first electron has tunneled. Furthermore, the different possible
intermediate states add {\it coherently} and the effective matrix element
$\hat{T}$ (which should be entered in golden rule formula (\ref{GoldenRule}))
between initial and final state becomes
\begin{equation}
\left\langle f|\hat{T}|i\right\rangle \approx \frac{1}{\epsilon
_{k}-\epsilon _{p}-E^{+}}+\frac{1}{\epsilon _{p^{\prime }}-\epsilon
_{k^{\prime }}-E^{-}}.  \label{T}
\end{equation}
The electrostatic energy differences between the initial and the two
possible intermediate states denoted by $E^{+}$ and $E^{-}$ correspond to
adding or removing an electron from the island, respectively. In order to
find the rate for transfer of an electronic charge from left to right one
must sum over all $k,k^{\prime },p,p^{\prime }$, and furthermore include
appropriate occupation probabilities. At zero temperature these integrals
can be performed \cite{aver89}.

At low temperature and small bias $k_{B}T,eV\ll E^{\pm }$ one can
approximate the intermediate state energies in denominators in Eq.\ \ref{T}
by the electrostatic energy differences, because energy of the electron-hole
pairs is limited by temperature and bias voltage. In this case one finds for
the cotunneling rate through the device from left to right \cite{aver90}
\begin{equation}
\Gamma _{{\rm cotun}}(L\rightarrow R)\approx \frac{G_{L}G_{R}\hbar }{12\pi
e^{3}}\left( \frac{1}{E^{+}}+\frac{1}{E^{-}}\right) ^{2}\left[ (2\pi
k_{B}T)^{2}+(eV)^{2}\right] \frac{V}{1-\exp (-eV/k_{B}T)},  \label{cotunlow}
\end{equation}
where $V$ is voltage drop across the device.

The approximate formulae Eq.\ (\ref{cotunlow}) and the zero temperature
result has been tested in several experiments \cite
{eile92,pasq93,paul94,mats95} and very good agreement was found. However,
outside the range of validity for the approximations, i.e. in the cross-over
between the cotunneling dominated and sequential tunneling dominated regimes
the theory \cite{aver89,aver90} gives unphysical divergences in the
cotunneling rate and needs to be modified.

To repair the possible divergence in (\ref{T}) several authors \cite
{pasq93,koro92,naza93b,aver94,flen97} have included a life time broadening
due to tunneling in and out of the island . The life time broadening results
in a imaginary term in the denominators of Eq.\ (\ref{T}) and in this way a
gradual cross-over between the cotunneling and the sequential tunneling
limits is obtained. Comparison with experimental data has shown that good
agreement is obtained for not too small tunnel resistances \cite
{pasq93,geer94}. In passing we note that recent work by K\"{o}nig, Schoeller
and Sch\"{o}n \cite{koni97} has shown that a thorough second order
perturbation theory, which does not suffer the divergence problem, gives
very good agreement with experiment even for rather small tunnel resistances
\cite{joye97}. Unfortunately, these approaches are complicated and it is not
clear at this point whether generalization to the cotunneling through many
junctions can be made sufficiently effectively for practical numerical
analysis of multi-junction devices. In order to have a simple way of
evaluating the cotunneling contributions, Jensen and Martinis \cite{jens92}
have proposed the following approximation scheme. They evaluated the
integration over intermediate states by distributing energy difference
between initial and final states equally among the electron-hole pairs. This
means that for the matrix element in Eq. (\ref{T}), $\epsilon _{k}-\epsilon
_{p}=\epsilon _{p^{\prime }}-\epsilon _{k^{\prime }}=(E_{i}-E_{f})/2.$
Fonseca et al. improved this approximation by shifting the lowest order
tunneling rate in Eq. (\ref{Gamma}), to lower energies. By doing this these
authors obtained a smaller difference between the approximate tunneling rate
and the one obtained from the more correct higher order approximations \
cite{pasq93,koro92,naza93b,aver94,flen97,lafa93}.

The expression for several consecutive cotunneling events was derived by
Averin and Odintsov \cite{aver89}. The Jensen and Martinis and the Fonseca et
al. approximations are easily generalized for the multiple cotunneling case
and they are used in the numerical simulations of pumps and turnstiles,
reviewed in Sec.\ \ref{sec:simulations}. We give here the final expression
for the a multiple cotunneling events of order $n$ \cite{jens92}, because it
is the dominant source of error at the lowest temperatures:
\begin{equation}
\Gamma _{{\rm cotun}}^{(n)}=\frac{2\pi }{\hbar }\left( \prod_{i=1}^{n}\frac{\hbar G_{i}}
{2\pi e^{2}}\right) S^{2}F_{n}(\Delta E_{n}),  \label{manycotun}
\end{equation}
where $G_{i}$ is the conductance of tunnel junction number $i$, $\Delta E$
is the energy difference between final and initial states, and
\begin{equation}
S=\sum_{{\rm permutations}}\left( \prod_{k=1}^{n-1}\frac{1}{\Delta
E_{k}-\Delta E/n}\right)  \label{eq:S}
\end{equation}
where the sum is over all permutations of cotunneling sequences going from
initial to final state, and finally
\begin{equation}
F_{n}(\Delta E)=\frac{\prod_{i=1}^{n-1}\left[ (2\pi k_{B}Ti)^{2}+(\Delta
E)^{2}\right] }{(2N-1)!}\frac{\Delta E}{\exp \left( \Delta /k_{B}T\right) -1}.
\end{equation}
From this expression we see that the higher order cotunneling events are
suppressed by the ratio of the junction conductances to the quantum
conductance $e^{2}/h$ to the power $n$. For typical junctions with
resistances in the 100 k$\Omega $ range the $n$ order cotunneling process
contains a factor $(0.04)^{n}$ due to the first product of Eq.\ (\ref
{manycotun}). However, in contrast to the thermally activated tunneling it
is not exponentially suppressed and the cotunneling leakage events will thus
dominate at low temperatures.

One way of suppressing cotunneling is to increase the number of Coulomb islands.
Another method which was suggested in Ref.\ \cite{odin92} is to increase the resistance
of the leads that couples to the device. The authors of ref.\ \cite{odin92} showed that for a
two junction circuit the cotunneling rate the cubic voltage dependence in
Eq.\ (\ref{cotunlow}) is replaced by $\Gamma_{\mathrm{cotun}} \sim V^{3+2z}$,
where $z=Re^2/h$ is the normalized resistance of the leads. This idea is now
being pursued experimentally\cite{zorin:private}

Finally, we mention the so-called {\it elastic }cotunneling, which means
that energy is not dissipated to electron hole pairs in the tunneling
process. Consequently, the electron states $p$ and $p^{\prime }$ in Fig.\
\ref{fig:cotun} are the same. In other words, it is ``the same electron''
that enters and leaves the island. The probability of elastic cotunneling is
proportional to the single particle level spacing \cite{aver90}.
Thus the process is generally very weak for metal islands.
However for semiconductor devices this effect indeed gives significant
contributions to the electron transport.

\subsection{Photon assisted tunneling}

\label{sec:photon}When a SET device is irradiated by an external
electromagnetic field, the potentials on the electrodes will oscillate.
Tunneling events which are otherwise not possible may now occur through
absorption of photons from the external source. Only small amounts of
electromagnetic radiation are needed to introduce considerable errors due to
photon assisted tunneling \cite{mart93}. Therefore careful attenuation of
the high frequency noise is necessary by means of shielding, thermally
anchoring all leads in the cryostat, and applying copper powder filters \cite
{mart87}, lithographically made miniature filter \cite{vion95} or lossy
coaxial cables \cite{zori95}.

When the device is driven by a periodic signal as in the pump or turnstile,
it is possible that the drive itself induces photon assisted cotunneling
events. This effect was considered in Ref. \cite{flen97}, but it is was
found that only at a frequency higher than several gigahertz this error
contribution becomes significant. It can thus be neglected for present days
pumps and turnstiles operating at typically 5-10 MHz.\smallskip

\subsection{Background charges}

\label{sec:background}

Noise introduced by a random motion of charges in the dielectric surrounding
(the substrate, the barriers of the tunnel junctions, etc.) of the
conducting islands is a serious problem in SET devices. This noise leads to
fluctuations of the island's polarization. Usually, the frequency dependence
of the noise spectral density is close to $1/f$, although sometimes it
exhibits an almost $1/f^{2}$ dependence in combination with a telegraph-like
time dependence \cite{verbrugh,bouchiat}. This switching noise results from
one or several two-level fluctuators with typical switching times ranging a
fraction of a second up to minutes. For most of the aluminium-based SET
devices, fabricated on various substrates (SiO$_{2}$, Si, Al$_{2}$O$_{3}$)
the intensity of the background charge noise at $10$ Hz is between $10^{-4}$
and $10^{-3}\,e/\sqrt{\text{Hz}}$ \cite{verbrugh,bouchiat,zimm92,zori96}.

However, in the devices fabricated with a stacked design the islands are
partly \cite{viss95,krup1,krup2} or entirely \cite{krup1,krup2} screened
from the substrate by an underneath gate or by an underneath counter
electrode. Such devices exhibit considerably less noise, down to the lowest
value measured so far, $2.5\times 10^{-5}$ $e/\sqrt{\text{Hz}}$ at $10$ Hz
\cite{krup1,krup2}. The dominating role of substrate in causing the
background charge noise is also proven by the mutual correlation observed in
the cross spectrum of the $1/f$ noise of two SET electrometers which were
closely positioned on the same substrate \cite{zori96,mann96}. Therefore, a
suitably chosen substrate and/or an appropriate geometry for SET devices
should reduce the background charge noise.

The effect of background charges, as discussed above, is not limited to time
dependent fluctuations. The dc offset charges \cite{mart94,kell97} can also
arise due to charges trapped inside the substrate and the barriers of the
tunnel junctions, and due to different work functions of dissimilar
electrode materials \cite{kuli75}.

\section{Numerical analysis of single electron devices}

\label{sec:simulations}

\subsection{Methods and tools for numerical analysis}

Analytical treatment of the single-electron transport is only possible for
circuits with few tunnel junctions such as the SET transistor, or symmetric
devices such as one-dimensional tunnel junction arrays. In the last few
years, several computer codes have been developed to analyze complex single
electron circuits numerically \cite
{Nanosimul,ChenMoses,fons95,fons96a,fons96b,SIMON}. Two main approaches have
been used. Firstly, Monte Carlo algorithms \cite
{Bakhvalov,Kirihara,Roy,ChenMoses,SIMON}, which average over various
realizations of stochastic SET events. Secondly, algorithms based on a
numerical solution of the master equation (\ref{masterequation}), which
describes the dynamics of SET circuit in terms of time-dependent
probabilities of various charge configurations
\cite{jens92,poth91,fons96a,fons96b,fons95,SIMON,aver93}.

The advantage of Monte Carlo routines is their relative simplicity and
robustness. This makes them convenient tools for the study of the dynamics
of single electron systems, for which rare events are not important. Another
feature of the Monte Carlo method is the trade-off between accuracy and
simulation time: one can quickly achieve approximate results for very large
circuits, which otherwise would not be possible to compute. An intrinsic
drawback of Monte Carlo algorithms is their inefficiency in the analysis of
rare events, simply because these rare events might not be encountered
during the simulation time. Specifically we mention the codes MOSES \cite
{ChenMoses} and SIMON \cite{SIMON} which are able to analyze arbitrary
circuits consisting of tunnel junctions, capacitances, resistors, and signal
sources. The latter code is convenient because of its graphical user
interface.

Reliable simulations of SET devices require realistic estimates of the
contributions of rare events. This can be achieved by starting from the
master equation (\ref{masterequation}), whose solution contains {\it complete}
 statistical information about the dynamics of the system including the
probabilities of the rare events. The problem with the master equation
method is the necessity to trace the time-evolution of a large number (up to
several thousands) of charge states. This makes the computation
time-consuming. The analysis is alleviated in the case of adiabatic devices
(like single electron pump) where the electron transfer occurs at low
energies compared to $E_{C}$ and the master equation involves a moderate
number of low energy charge states \cite{jens92,poth91,aver93}. In order to
use the master equation approach for general, non-adiabatic devices (like
turnstiles) the development of more sophisticated algorithms is required. In
the following section, we will briefly describe an effective algorithm of
this type that is based on the ideas of global analysis and the dynamical
choice of the basis of states \cite{fons95,fons96a,fons96b}.

Finally, we mention the attempt to combine the Monte Carlo approach with
solution of the master equation \cite{SIMON}. An essential assumption of
this method is that the system can stay in a rare state for a period much
shorter than the characteristic time of the simulation (e.g. the period of
an external signal in case of pumps). This assumption in fact means that the
errors due to rare events are analyzed {\it locally} in time. In the
following section it is show that in order to detect errors in the operation
of SET devices, the errors should be analyzed {\it globally}, throughout the
whole cycle of the device operation.\smallskip

\subsection{SENECA: a global approach to rare events}

In this section, we introduce SENECA (Single Electron NanoElectronic Circuit
Analyzer), a computer algorithm suitable for studies of the dynamics and
statistics of single electron systems consisting of arbitrary combinations
of small tunnel junctions, capacitances, resistors and voltage sources \cite
{fons95,fons96a,fons96b}. The method is based on the numerical solution of a
linear matrix equation for the vector of probabilities of various electric
charge states of the system, with iterative refining of the operational set
of states. The method is able to describe very small deviations from the
classical behavior of a system, caused by the physical error mechanisms
discussed in Sec.\ \ref{sec:errors} . Below, the main underlying ideas and
characteristics of the algorithm are briefly discussed.

{\it Global analysis of errors.} It must be emphasized that one should
distinguish between a rare tunneling event and the error in operation of a
device (i.e. deviations from Eq. (\ref{Ief})). Some rare tunneling events
bring the system to an unwanted charge state, which afterwards evolves
classically to a correct operational cycle. These rare events do not affect
the proper operation of the device. In order to separate those events from
the error introducing events, one should analyze the operation globally by
solving the master equation throughout the whole cycle.

{\it Clustering of the charge states according to their probability -
iterative approach. }For single electron devices of interest the number $M$
of islands ranges from 5 to 30. If we consider the charge configurations of
the islands with additional charges $0$ and $\pm e$, the number of states
can be roughly estimated as $3^{M}$. Nevertheless, when the temperature is
low and the tunnel conductances of the junctions are small, most of the
possible transitions (either those leading to an increase in the
electrostatic energy, or those of high cotunneling order) have very low
rates. As a consequence, a large number of charge states have very low
probabilities $P$ and may be ignored. Hence, an iterative approach is chosen
which considers for each iteration $n$ only the limited number of states for
which $P>P(n)$. The threshold $P(n)$ decreases rapidly with $n$.

{\it Dynamical choice of the basis of states. }Each iteration corresponds to
the solution of the master equation at the time interval of the operation
cycle. This interval is divided into enough time steps, so that the
variation of external parameters between steps can be neglected. At each
time step the algorithm first guesses the set of states whose probabilities
are expected to be large enough, $P>P(n)$, then solves the master equation
for this set, and finally filters out the states whose probabilities at the
end of the step turned out to be low, $P<P(n)$. In this way the minimum
possible basis of states is kept throughout the computation.

\subsection{Results of simulations}

\label{sec:results}
\begin{figure}[tbp]
\setlength{\unitlength}{1.0cm}
\begin{picture}(14,7)
\put(0.,1.0){\epsfxsize=7cm\epsfbox{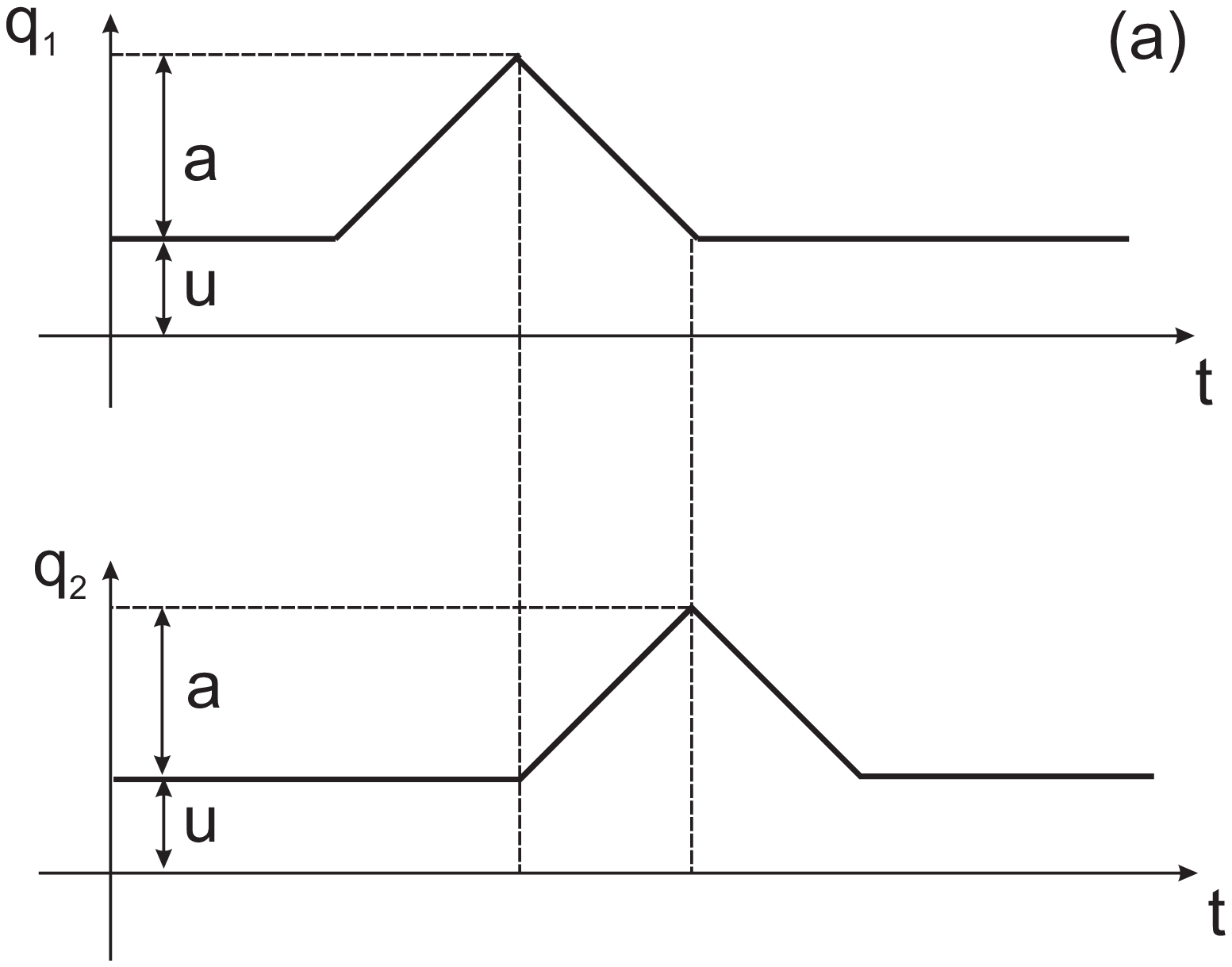}}
\put(8.0,2.0){\epsfxsize=7cm\epsfbox{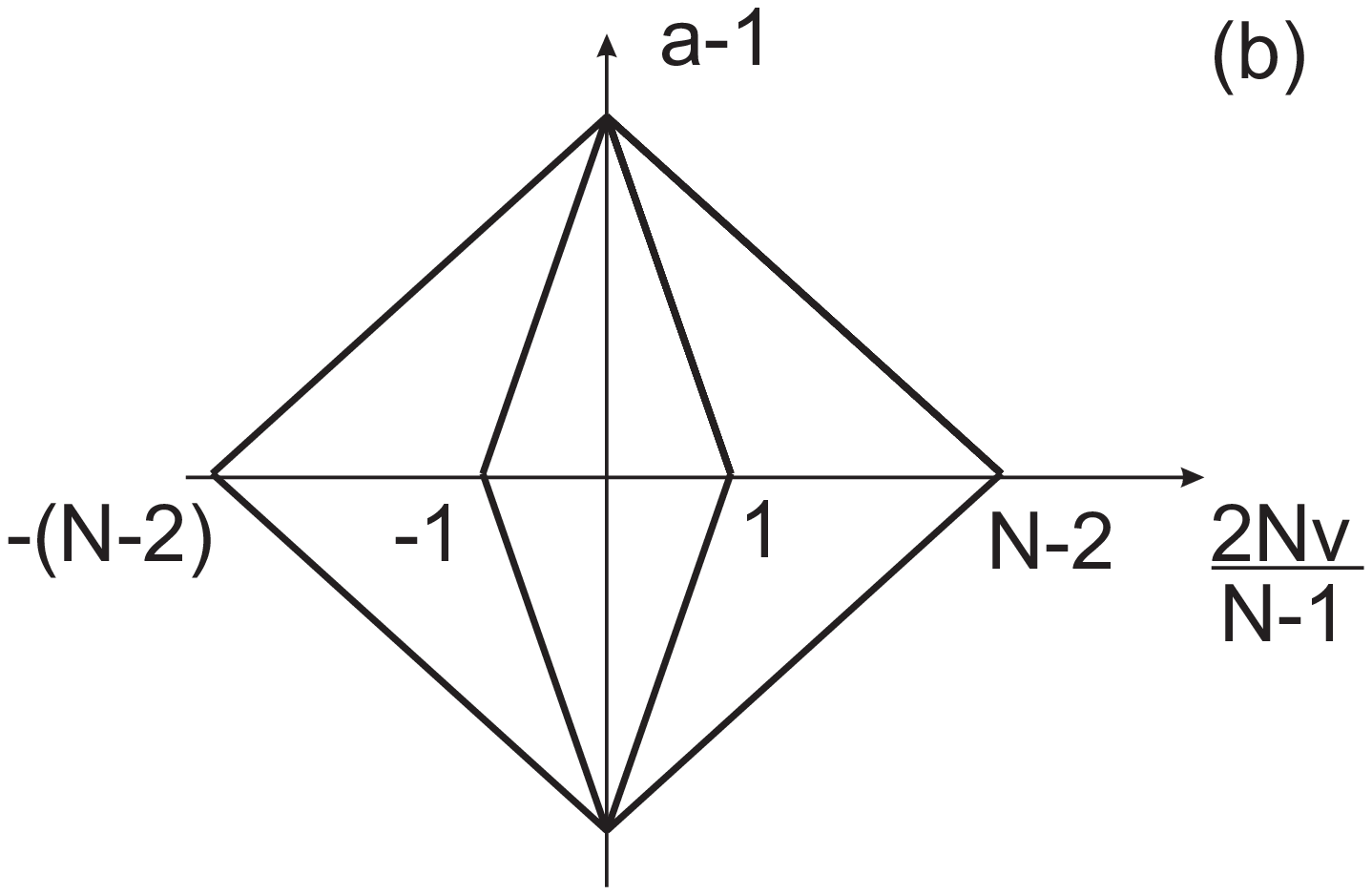}}
\end{picture}
\vspace{-0.5cm}
\caption{(a) The shape of the modulation of the gate voltages $V_{Gi}$
($q_{i}=C_{Gi}V_{Gi}/e$) in the pump device. (b) The region of correct
classical pump operation in $(a,v)$ plane (see text). The dashed area is the
region where the lowest order of allowed co-tunneling transitions is $N-1$.
From Ref. \protect\cite{aver93}.}
\label{fig:pumpdiagr}
\end{figure}

{\it Pump.} First we consider an $N$-junction pump driven
by triangular voltage pulses with amplitude $ae/C_{Gi}$ and offset
$ue/C_{Gi} $ applied to the gates as shown in Fig.\ \ref{fig:pumpdiagr}(a).
The gate capacitances are assumed to be small: $C_{Gi}\ll C$. If the pump
works properly, an electron tunnels sequentially through all junctions of
the array following the propagation of the pulse. The region of correct
classical operation (disregarding any rare processes) can be evaluated from
the condition that tunneling of an electron occurs only in an ''active''
junction biased by a voltage pulse, while the other junctions are closed. By
choosing $u$ equal to $(1-a)/N$ one maximizes the region of correct
classical operation (Fig.\ \ref{fig:pumpdiagr}(b),
\begin{equation}
|v|<\frac{(N-1)(N-2)}{2N}\min [a,(2-a)]\;,  \label{cl_oper}
\end{equation}
with $v$ being the bias voltage in units of $e/C$.

The closer the parameters $a$ and $v$ are chosen to the center of the region
(\ref{cl_oper}), the smaller the error due to rare events. In particular,
the order $n$ of cotunneling processes, which, according to Eq. (\ref
{manycotun}) primarily determines the intensity of cotunneling, increases
from $2$ near the inner borders of this region to $N-1$ in its central part
(dashed area in Fig. \ref{fig:pumpdiagr}),
\begin{equation}
|v|<\frac{(N-1)}{2N}\min [a,(2-a)]\;,  \label{central_part}
\end{equation}
in which the best performance of the device occurs. In the region (\ref
{central_part}) the deviation $\Delta I=I-ef$ \ of the current $I$ from its
ideal value $ef$ is determined by two competing cotunneling processes, shown
Fig. \ref{fig:5pump}. In one process instead of regular tunneling through an
active junction (corresponding, for example, to the transition
$(0,0)\rightarrow (0,1)$ in Fig. \ref{fig:2pump}), the cotunneling occurs
through all the other $N-1$ junctions in the opposite direction
($(0,0)\rightarrow (1,0)\rightarrow (0,1)$, $(1,0)$ being the intermediate
state, see Sec. \ref{sec:cotunneling}). In this process the electron charge
is transferred through the device in the direction opposite to the current.
The competing cotunneling process starts {\em after} the regular tunneling
event and transfers an electron charge in the direction of the current
($(0,1)\rightarrow (1,0)\rightarrow (0,0)$ in Fig. \ref{fig:2pump}).

In the region of optimal operation (\ref{central_part}) the error $\Delta I$
in the current can be evaluated analytically \cite{jens92,aver93}. The
performance of the device can be characterized by the slope of the current
platoe, $G=d(\Delta I)/dV$ at $V=0$. Being multiplied with the size $e/C$ of
the optimal region (\ref{central_part}) this gives characteristic scale of
the current deviation $\Delta I^{*}$, or accuracy of the device,
$\varepsilon =\Delta I^{*}/ef=G/fC$.

We consider first the case of a slow drive, $k_{B}T\gg eV_{0}\equiv
e^{2}[(N-1)\dot{q}_{i}R_{T}/NC]^{1/2}$, $\dot{q}_{i}\simeq Nf$, $R_{T}$
being the tunnel resistance of the junctions. In this adiabatic regime the
probabilites to find an electron on the left and right electrodes of an
active junction correspond to equilibrium conditions at any time. The
deviation $|\Delta I|$ of the current due to cotunneling increases with
temperature and does not depend on driving frequency. The accuracy is given
by \cite{aver93,jens92},
\begin{equation}
\varepsilon \simeq \frac{Ne^{2}R_{T}}{\hbar CV_{0}^{2}} \left( \frac{\pi
\hbar}{2e^{2}R_{T}} \right)^{N-1} S^2 (k_{B}T)^{2N-3} \;,  \label{acc_highT}
\end{equation}
where the factor $S$ (given by Eq. \ref{eq:S} with $\Delta E = 0$), can be
found explicitly in the center ($a=1$, $u=0$) of the optimal region (\ref
{central_part}),
\begin{equation}
S= \left( \frac{2NC}{e^{2}} \right)^{N-2} \frac{N-1}{(N-2)!} \;.
\label{S_pump}
\end{equation}

In the opposite case, $k_{B}T\ll eV_{0}$ the tunneling through an active
junction occurs non-adiabatically (effectively only a single tunneling event
occurs during the gate voltage pulse); the deviation $|\Delta I|$ of the
current due to cotunneling increases with increasing driving frequency and
does not depend on temperature \cite{aver93,iwas95}. The accuracy can be
evaluated as \cite{aver93,jens92},
\begin{equation}
\varepsilon \simeq \frac{e}{CV_{0}} \frac{NS^{2}}{(2N-3)!} \left( \frac{
\hbar V_{0}^{2}}{\pi R_{T}} \right)^{N-2} \;.  \label{acc_lowT}
\end{equation}
Hence, for a given temperature, the optimal driving frequency corresponds to
the crossover point, where the characteristic energy $eV_{0}$ of electron
tunneling is equal to the energy $k_{B}T$ of thermal fluctuations.

So far we have not considered the rare events due to thermally activated
classical tunneling whose probability decreases exponentially with
temperature, $P_{th}\sim \exp (-\Delta E/k_{B}T)$, $\Delta E\simeq E_{C}$.
The analysis \cite{jens92} has shown that thermal activation of classical
tunneling can be neglected compared to the cotunneling at temperature of the
order of or smaller than $0.025E_{C}$, which is usually the case in
experiments.

\begin{figure}[tbp]
\begin{picture}(12,12)
\put(1.5,-2){\epsfxsize=12cm\epsfbox{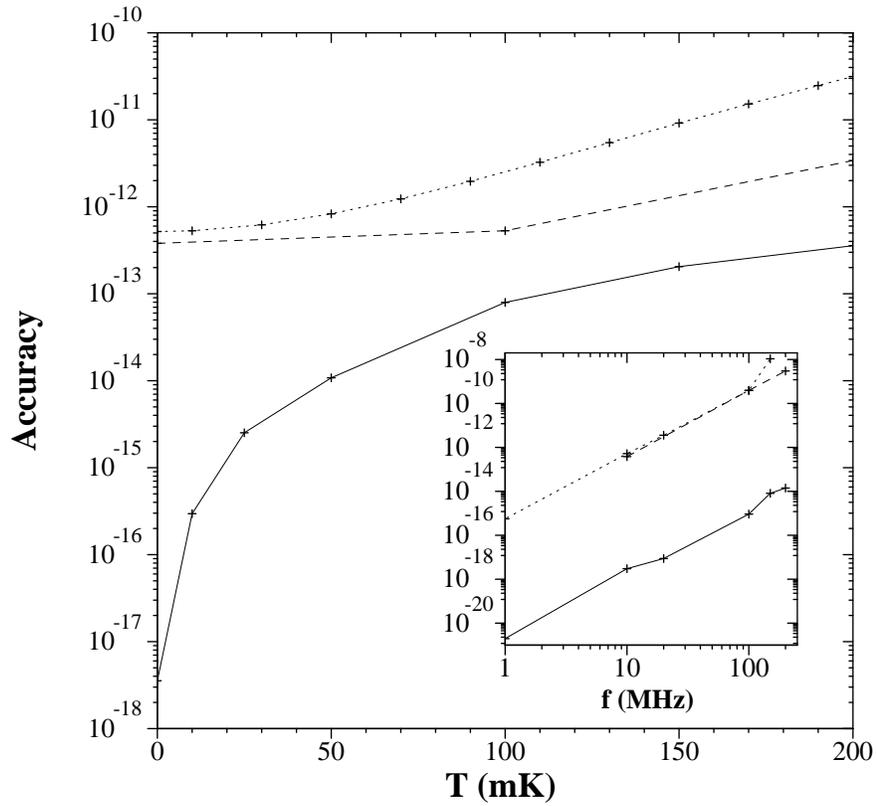}}
\end{picture}
\par
\vspace{0cm}
\caption{Accuracy of the 5-junction pump driven by optimized step-like
waveform (solid line), triangular waveform with optimized tunnel resistances
(dashed line), and triangular waveform with fixed tunnel resistances
$R_{T}=300$ k$\Omega $ (dotted line). The insert shows the accuracy as a
function of driving frequency at $T=0$. From Ref. \protect\cite{fons96a}. }
\label{fig:senecapump}
\end{figure}

The early estimates \cite{jens92,aver93} of the abovementioned factors
confirmed by more recent simulations using SENECA (see Fig.\ \ref
{fig:senecapump}) \cite{fons96a,fons96b} predict that already the 5-junction
pump with typical experimental parameters, $R_{T}=300$ k$\Omega $, $C=0.1$
fF, $f=10$ MHz, $T=100$ mK should have metrological accuracy $\varepsilon
\simeq 2\cdot 10^{-12}$. It should be noted that both the cotunneling and
thermally activated tunneling can be effectively suppressed further by
increasing the number of the junctions and decreasing the junction size
(and/or the operation temperature).

Another way of improvement of the pump operation is the optimization of the
driving voltage pulse shape \cite{fons96a,fons96b}. The idea is based on the
consideration of the energy dependences of the classical tunneling, Eq. (\ref
{Gamma}), $\Gamma _{cl}\propto \Delta E$ and cotunneling rates $\Gamma
_{\cot }\propto \Delta E^{2N-1}$, Eq. (\ref{manycotun}). To minimize the
ratio $\Gamma _{\cot }/\Gamma _{cl}$ one has to decrease $\Delta E$ during
the ''active'' portion of the time interval when the electron transfer
between the islands occurs. When the probability of the transfer is close to
unity, one increases $\Delta E$ to ensure the full transfer of an electron
during the ''passive'' portion of the interval. Being optimized by SENECA
algorithm, this two-step drive improves the calculated accuracy of
5-junction pump (Fig.\ \ref{fig:senecapump}) by a factor of $30$ for
$k_{B}T\simeq 100$ mK and other parameters given above (the improvement
enhances at lower temperatures).

\begin{figure}[tbp]
\begin{picture}(8,8)
\put(3.5,0){\epsfxsize=8cm\epsfbox{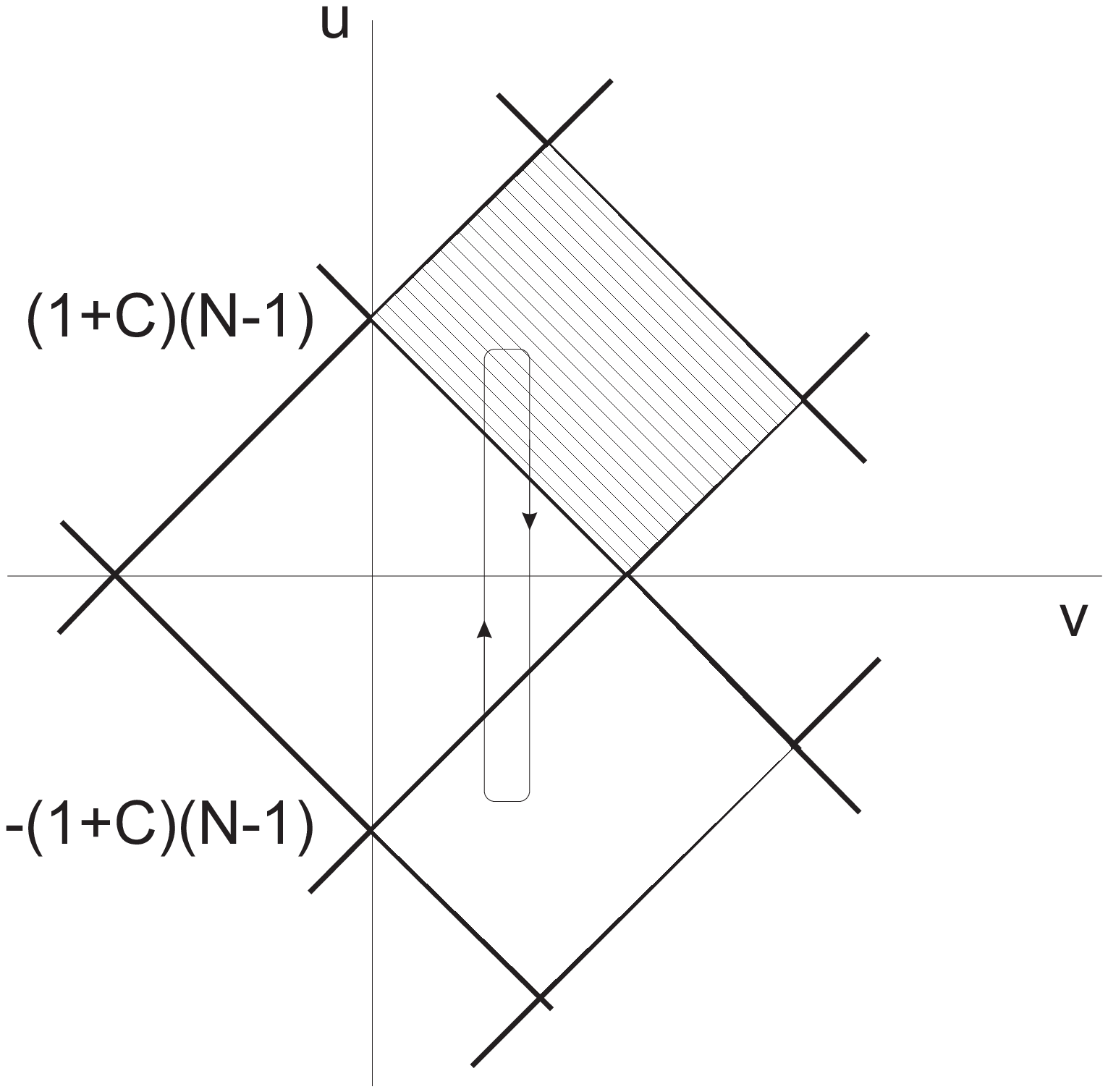}}
\end{picture}
\vspace{0.5cm}
\caption{The diagram of the turnstile operation in the
$v - u$ plane, $v=(2C+NC_{G})V/e$,
$u=N(2C_{G}V_{G}/e-1)$. In the upper dashed region an electron is pulled in
the turnstile, in the lower dashed region it is pushed out. The central
square corresponds to the Coulomb blockade state where all classical
transitions are suppressed. From Ref. \protect\cite{aver93}.}
\label{fig:turnstoperation}
\end{figure}

{\it Turnstile.} The diagram of correct classical operation of the turnstile
with $2N$ junctions is shown in Fig.\ \ref{fig:turnstoperation} (see also
Fig.\ \ref{fig:turnstile}). Being driven by the gate voltage alternating
between the two dashed regions in Fig.\ \ref{fig:turnstoperation} the device
transfers one electron per cycle of the gate voltage. The regions are
defined by \cite{aver93}
\begin{eqnarray}
u+v &>&(1+c)(N-1),\;u-v<(1+c)(N-1)  \label{turn_cl_oper1} \\
u+v &<&(3+c)N-1-c,\;v-u<(1+c)(N-1)  \label{turn_cl_oper2}
\end{eqnarray}
and similar pair of equations with $u\rightarrow -u$. Here
$u=(2C_{g}V_{g}/e-1)N$, $v=(2C+NC_{g})V/e$, $c=NC_{g}/C$.

\begin{figure}[tbp]
\begin{picture}(12,12)
\put(1.5,0){\epsfxsize=12cm\epsfbox{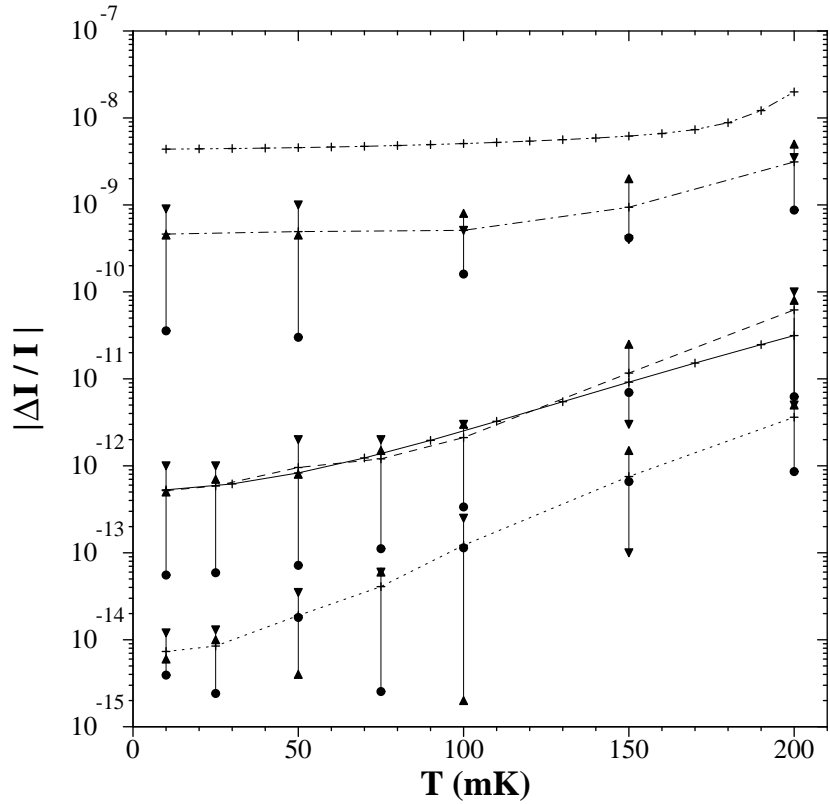}}
\end{picture}
\caption{Deviation of the current from the quantized value in the unbiased
(solid line) and biased (dotted line) 5-junction pump , unbiased
(dot-dot-dashed line) and biased (dashed line) 6-junction hybrid pump (all driven
by a triangular waveform), and 8-junction turnstile driven by a step-like
waveform (dot-dashed line) at $f=10$ MHz as functions of temperature.
Circles: current deviation at the inflection point $V_{0}$, triangles up
(down): current deviation at $V_{0}(1 \pm 0.1)$. From Ref. \protect\cite
{fons96b}. }
\label{fig:senecacomp}
\end{figure}

Analysis \cite{aver93} shows that the order of cotunneling transitions
reaches its maximum $n_{0}=0.8N$ ($N\gg 1)$ for $c<2/3$ and the voltages
near the threshold of correct classical operation ($u_{high}+v=(1+c)(N-1),$
$u_{low}-v=-(1+c)(N-1)$, see Fig.\ \ref{fig:turnstoperation}). This happens
at the optimal voltage $V_{opt}=(1+c)(n_{0}-1)e/[2(2+c)C]$. Careful
numerical optimization \cite{fons96a,fons96b} of the accuracy of 6 and 8
-junction turnstiles as function of the gate capacitance $c$ and the
amplitude $u_{high}=-u_{low}$ of the gate voltage has shown that the optimal
gate capacitance corresponds to $c\approx 2/3.$ The calculated uncertainty
of 8-junction turnstile with optimized parameters was better than $1$ part
in $10^{9}$ for $R_{T}=300$ k$\Omega $, $C=0.1$ fF, $f=10$ MHz, and $T=100$
mK (see Fig.\ \ref{fig:senecacomp}).

{\it Hybrid pump-turnstile.} In order to reduce the number of parameters (dc
gate voltages, amplitudes and phases of RF gate voltages, bias voltage)
controlling the operation of a pump device, it has been proposed to apply RF
signal to every second gate of the pump. The calculated uncertainty of a
properly biased pump-turnstile with 6 junctions \cite{fons96a,fons96b} was
close to the uncertainty of unbiased pump ($2$ parts in $10^{12}$ for the
parameters listed above), but deteriorated to $5$ parts in $10^{9}$ for
unbiased device, see Fig.\ \ref{fig:senecacomp}.

\section{Comparison of theory and experiments}

\label{sec:compare}

\subsection{Discrepancies between measured and predicted errors}

\label{sec:discrep} The experiments reviewed in the Sec.\ \ref{sec:exp} have
demonstrated an error per pumped electron of the order of $1$ part in $10^{8}$
for pump-like SET devices. Up to very recently, the development followed
the most obvious path, namely the improvement of the accuracy by increasing
the number of tunnel junctions (which suppresses cotunneling) and the
reduction of their size (which enhances the Coulomb blockade). In particular
the 7-junction pump has been very well characterized \cite{kell98} with
respect to the capacitances of the individual junctions, the offset charges
at the islands, as well as the electron temperature.

Substantial efforts have been made \cite{mart94,kell98} to compare the
observed accuracy of the pump device with theoretical predictions and to
identify unforseen error mechanisms. Some details of the experiments are
given in Sec. \ref{sec:NIST}.

For the 5-junction pump \cite{mart94} the error per pumped electron at 40 mK
has been observed to be exponentially dependent on pump frequency for
frequencies higher than 5 MHz. This behaviour is expected when cycle missing
is the dominant error mechanism (see \ref{sec:cycle}). However, for
frequencies smaller than 5 MHz, the observed error per pumped electron at 40
mK saturates at a level of $1$ part in $10^{6}$. This saturation, when
ascribed to thermally activated processes and cotunneling, occurs at a value
that is $10^{8}$ times larger than
expected. In order to explain this very large discrepancy Martinis {\it et
al.} \cite{mart94} suggested that the excess error is caused by an
additional source of energy, e.g. by leakage of thermal noise generated at 4
K or by energy released by charge traps in the substrate.

For the 7-juction pump \cite{kell96}, the measured error per pumped electron
at 140 mK, when the dominant error mechanism is thermally activated
tunneling, has been found to be $1$ part in $10^{6}$. This is in reasonable
agreement with the theoretical estimate of $1$ part in $10^{7}$. However, at
40 mK, when the dominant error mechansm is cotunneling, the measured error
per electron amounts to $1$ part in $10^{8}$, which is in sharp contrast to
the theoretically predicted value of $1$ part in $10^{20}$. Since the error
per electron is constant and only weakly dependent on the wait time between
the transfer of individual electrons, the errors are concluded to occur
predominantly during the active pumping cycle. Therefore, the error
mechanisms in the pump mode and hold mode are likely to be different.

Despite the fact that this 7-junction pump exhibits a much smaller error per
electron than the 5-junction pump, this improvement is not as drastic as
expected from theory, especially in the temperature range below 100 mK.

\subsection{Possible sources of additional error mechanisms}

\label{sec:additional} As mentioned in the previous section, additional
error sources can exist due to 4 K radiation or trapped charges in the
substrate. Usually, in order to prevent 4 K radiation from reaching the
device, special microwave filters are placed in series with every electrical
lead. In addition, the device is mounted in an electromagnetically shielded
box. Surprisingly, it was shown \cite{kell98} that removal of the
filters and the shield does not necessarily affect the error rate of the
7-junction pump. This hints at a strong internal low-temperature source of
spurious tunneling events, which is much stronger than the external noise.
It is also possible that photons are generated by non-equilibrium charge
fluctuations in the substrate and/or the tunnel barriers (see Sec.\ \ref
{sec:photon} and Ref. \cite{baue93}). Such an intrinsic photonic source
cannot be filtered, and may present a fundamental problem for metrological
and digital application of SET devices.

Note that it can be difficult to separate the effect of high frequency
photons due to the charge fluctuations from the effect of low frequency
noise due to the drift of the offset charges. However, in the experiments
reported by Ref. \cite{kell98} the comparison with theory was made when the
optimal background charge compensation was stable. Often the devices were
kept at millikelvin temperatures for several weeks, after which movement of
background charges was very small and adjustments of the dc voltages on the
gate lines was only necessary with time intervals of tens of hours.

\section{Alternative designs for single-electron current standards}
\label{sec:alternatives}
The main disadvantage of the discussed current standards based on normal SET
junctions is their relatively low operation frequencies of a few megahertz,
which correspond to a current level of only a few picoampere. This
limitation is caused by the necessity to have large tunnel resistances for
the normal junctions $R_{T}\gg R_{Q}\equiv h/e^{2}$ in order to fulfill the
condition of the charge quantization. The limitation in frequency may
however be circumvented in several ways, as will be discussed in this
section.

\subsection{Surface acoustic waves as carriers of single electrons}

\begin{figure}[tbp]
\epsfxsize=14cm\epsfbox{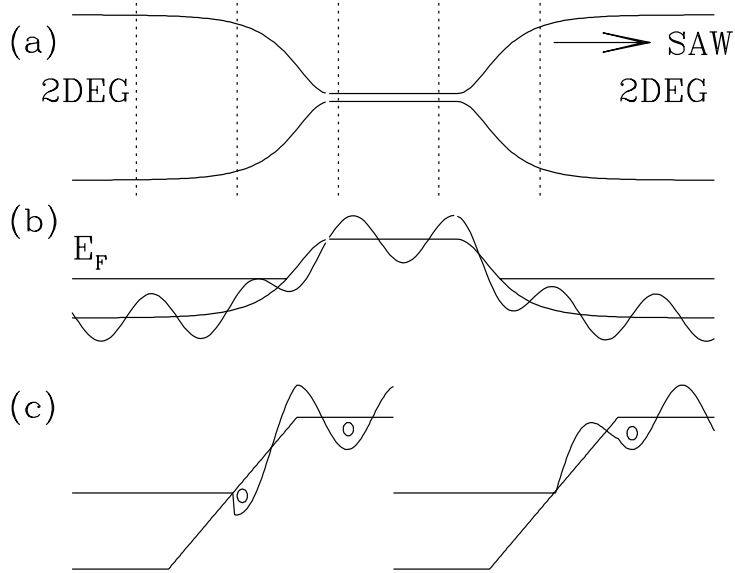} \vspace{-4cm}
\caption{ Schematic outline of a SETSAW device. (a) shows the two 2DEGs
separated by a narrow constriction. Propagating SAWs are indicated by dashed
lines. (b) illustrates the potential landscape along the center of the device.
In the constriction region a potential barrier is present. The SAW induces a
propagating modulation of the potential. In the 2DEG region the modulated
potential induced by the SAW is screened and the resulting potential is
shown in (c) for two different times. In the left part of the figure an
electron scooped up by a minimum in the potential after which it is moved upward
into the constriction region (right part of figure). }
\label{fig:sawdevice}
\end{figure}

The interaction of surface acoustic waves (SAW's) with a two-dimensional
electron gas (2DEG), present at the interface in a GaAs-AlGaAs
heterostructure, can be applied to generate an electrical current that is
determined by the frequency of the SAW. In Fig. \ref{fig:sawdevice} the
working principle of the so-called SETSAW device is outlined. Due to the
piezoelectric effect the SAW creates a propagating modulation of the
electrostatic potential through the 2DEG. Normally this modulation is
screened by the electrons in the 2DEG. By using a split-gate technique a
quasi-one-dimensional channel is formed in the 2DEG. By applying a
sufficiently negative voltage to the split gate the electrostatic potential
in the constriction is increased and the electron density can be reduced to
zero (pinch-off regime). In this case a propagating modulation in the
channel is no longer screened. For a sufficiently strong amplitude of the
SAW the quasi-one-dimensional electron gas in the constriction is split
into a sequence of small islands, or quantum dots, moving with the velocity
of the SAW.

Recently it has been shown that such a technique can be used to transport an
integer number of electrons per cycle of the SAW frequency \cite{shil96b,taly97,shil96a}.
The number $N$ of electrons per dot determines the
dc current $I=Nef$ through the constriction. The experiments have
demonstrated the quantization of the dc current at the values corresponding
to integer $N=1...4$. The robustness of the current quantization with
respect to a range of parameters (gate voltage, bias voltage, amplitude of
SAW) implies that the number of electrons in each quantum dot is a
deterministic rather than stochastic quantity.

The quantized current can only be observed deep inside the pinch-off regime.
The concentration $n=N/\lambda _{SAW}$ of electrons in the pinch-off region,
where $\lambda _{SAW}\approx 1 \mu $m is the wavelength of the SAW, is much
smaller than the concentration $n_{0}=4/\lambda _{F}$ of electrons in the
2DEG, where $\lambda _{F}\approx 50$ nm is the Fermi wavelength. For this
reason, the electrons in the constriction do not screen the electrostatic
potential induced by the SAW. Moreover, the kinetic energy of electrons in
moving quantum dots is very small, so that the Coulomb interaction is the
dominant energy scale. The quantum dot that has just been formed near the
entrance of the constriction can be considered as an electron box connected
to the 2DEG reservoir. The Coulomb repulsion determines the number of
electrons in the dot so that the free energy of the system is minimized. The
number of electrons in the dot can be controlled by the gate voltage and by
the amplitude of SAW (the increase of the SAW amplitude makes the potential
minima deeper which has a similar effect as an increase of the gate voltage).

Due to the substantial difference between the velocity of the SAW,
$v_{SAW}\simeq 3\times 10^{3}$ m/s, and the Fermi velocity, $v_{F}\simeq
10^{5}$ m/s, the formation of new quantum dots occurs adiabatically. In
contrast to a conventional electron box, the barrier between the quantum dot
and the reservoir varies in time so that its transparency decreases by many
orders of magnitude while electrons fill the quantum dot. The dynamics of
this process is yet to be understood. An important parameter that determines
the maximum operation frequency seems to be the time dependence of the
transparency (or conductance) of the barrier that separates the most
recently filled dot from the 2DEG.

The experimental observation of quantized currents at frequencies around 3
GHz suggests that electron transfer occurs through a low energy barrier,
although the accuracy of the device may be determined by residual thermal or
quantum transport through a high energy barrier. Finally we note that
another physical mechanism which might negatively affect the accuracy -
impurity scattering of the electrons in the moving quantum dots - becomes
less effective with increasing driving frequency \cite{taly97}.

\subsection{Phase locking of coherent Cooper pair tunneling}

\label{sec:cooper}
A current standard based on the tunneling of single
Cooper pairs may be an interesting alternative for the single-electron pump.
By using the Coulomb blockade state for Cooper pairs, one can take advantage
of the coherence which is intrinsic to the superconducting state. However,
achieving the Coulomb blockade for Cooper pairs is very difficult since the
effect can be masked by either the onset of a supercurrent or by unwanted
quasiparticle tunneling. Quasiparticle tunneling is presumably the main
cause of the poor performance of Cooper pair pumps \cite{geer90,geer91}.

Recently the blockade for Cooper pair tunneling was demonstrated in
one-dimensional series arrays of small-capacitance tunnel junctions with
tunable Josephson coupling \cite{havi96}. In this system the charge is
transferred by Cooper pairs {\it solitons \thinspace }which can be
considered as the dual analog of the vortices in a one-dimensional parallel
Josephson array. In particular, the Coulomb blockade voltage corresponds to
the threshold voltage at which Cooper pair solitons enter the array. In the
absence of dissipation, the Cooper pair solitons are expected to propagate
coherently through uniform arrays \cite{odin94}. A word of caution is needed
here. In realistic arrays the propagation of Cooper pair solitons might be
hindered due to the pinning by offset charges. The increase of the Josephson
energy effectively reduces this pinning. On the other hand, at $E_{J}\simeq
E_{C}$ the onset of superconductivity occurs in the array, which destroys
the Cooper pair solitons.

\subsection{Electron counting by a high-speed electrometer}

\label{sec:counter} The single-electron current standards discussed so far
are based on the {\it active} manipulation of electrons or Cooper pairs, in
other words current generation techniques. A tempting alternative is to try
to {\it passively} count individual charge quanta passing through a voltage
biased one-dimensional array of tunnel junctions using a high-speed
electrometer coupled to one of the islands of the array.

Due to correlations in the single-electron tunneling events, solitons of
charge $e$ (or $2e$) will be running through the array. Mutual repulsion
keeps the moving solitons at a certain average distance, so that the system
of solitons is robust with respect to the random offset charges of the
islands. Likharev et al. \cite{Bakhvalov} have calculated the time evolution
of the electric charge at an island of the array. In the frequency domain
narrow peaks show up, positioned at frequency $f=<I>/e$, where $<I>$ is the
average current. If it possible to monitor these charge oscillations in
time, one can actually realise an electron counter.

The main problem for the realisation of such an electron counter is to
develop a charge detector with a high charge sensitivity in combination with
a large bandwidth. Recently a promising step forward \cite{scho98} has been
made through the development of a radio frequency single-electron transistor
(RF-SET). The working principle has already been described in Sec. \ref
{sec:appl}. Possible error mechanisms in the electron counter based on
RF-SET have yet to be analyzed.

\section{Conclusions}

We have reviewed the status of the understanding of
single-electron devices in metrology. By pumping single electrons one can
generate a current which is determined by the pump frequency and the
fundamental electron charge. There are two possible routes for utilization
of this idea. One route is to devise a high precision {\it current standard}
based on the pumping principle. This would also allow for a comparison with
a current measured on the basis of a Josephson voltage standard and a
quantum Hall resistance standard. The direct comparison of the two currents
would give a check of the internal consistency of the quantum standards.
There are several technical difficulties in this, one of the main problems
being the small currents (presently a few picoamperes) generated by
single-electron pumps.

The second possible route is to realize a new {\it capacitance standard},
circumventing the cumbersome use of the calculable capacitor. The idea is to
pump a known charge onto a capacitor and then to measure the resulting
voltage. Note that this is a dc capacitance standard. In order to link its
value to that of the calculable capacitor, which operates at 1592 Hz, more
knowledge is required about the frequency dependence of these standards.

The state of the art 7-junction single-electron pump has an intrinsic error
per pumped electron of 1 part in $10^{8}$ \cite{kell97}, while generating a
picoampere current. This approaches the level of metrological accuracy
currently obtained for Josephson voltage standards and quantum Hall
standards. However, the measured error rate is many orders of magnitude
larger than that predicted by theory.

Although the quantum Hall effect is also not completely understood, it is an
inherent quantum effect, whereas single-electron current generators are
devices that have to be designed in order to {\it suppress} quantum
mechanical effects (e.g. cotunneling). Furthermore, the quantized Hall
resistance has been shown to be independent of material parameters, while
single-electron generators are very sensitive to the design layout and
choice of device parameters. For these reasons a detailed understanding of
the physical mechanisms of errors is essential for the use in metrology, and
future research is needed in this direction. Therefore, at this stage, the
single-electron pumps cannot be considered as being at a similar level as
the Josephson voltage standard or the quantum Hall standard, neither
experimentally nor theoretically.

In this review, we have mainly concentrated on conventional tunnel junction
SET devices such as pumps and turnstiles. Although the present accuracy is
still not sufficient for all metrological applications, the ongoing
technical developments justify the expectation of considerable improvements in
device performance. For future designs several issues are important. For
example to try waveforms which have been predicted to be much better than
the triangular shape. Secondly, to minimize noise from charge traps in the
substrate by alternative designs or materials. Also, the complexity of the
design of turnstiles and pumps may be reduced by introducing high-ohmic
microresistors in the bias leads thus suppressing cotunneling.
In this way one could obtain an error per
transferred electron equal to that of a 7-junction pump, but with less
tunnel junctions. It makes the operation of the device easier and makes the
device more robust.

In addition, there are promising candidates for alternative single-electron
current standards. One example is the SAW device, another example is a
single-electron counter based on a high-speed SET transistor.

A large worldwide research is evolving in the field of manipulation of
single electron charges. The efforts give enough confidence in a future
realisation of a high precision SET current standard.

\section{Acknowledgments}

We are indebted to A. Zorin and D. Haviland for their input to Secs. \ref
{sec:background} and \ref{sec:cooper} and to M. Keller for suggestions and
careful reading of the manuscript. We thank J. Martinis, H. D. Jensen, and
A. Hartland for discussions. This work was supported by the EU (SETamp
project, contract number SMT4-CT96-2049).

\end{document}